\documentclass[prb,floatfix,showpacs,superscriptaddress,longbibliography,twocolumn]{revtex4-1}

\usepackage[USenglish]{babel}

\usepackage{amsmath,amssymb,amsfonts,mathrsfs}

\usepackage{bm}

\usepackage{epsfig}
\usepackage{subfigure}

\usepackage[colorlinks=true,linkcolor=blue,citecolor=red]{hyperref}

\newcommand{\be}{\begin{equation}}
\newcommand{\ee}{\end{equation}}
\newcommand{\beS}{\begin{equation*}}
\newcommand{\eeS}{\end{equation*}}
\newcommand{\bea}{\begin{eqnarray}}
\newcommand{\eea}{\end{eqnarray}}
\newcommand{\ba}{\begin{eqnarray*}}
\newcommand{\ea}{\end{eqnarray*}}
\newenvironment{eqs}%
{\begin{equation} \begin{aligned}}%
{\end{aligned} \end{equation} }
\newcommand{\bal}{\begin{eqs}}
\newcommand{\eal}{\end{eqs}}
{\begin{equation} \begin{split}}%
{\end{split} \end{equation}}
\newcommand{\bas}{\begin{eqs}}
\newcommand{\eas}{\end{eqs}}
\newcommand{\Tr}{\text{Tr}}
\newcommand{\bw}{\begin{widetext}}
\newcommand{\ew}{\end{widetext}}

\begin{document}

\title{Transport through a magnetic impurity: a slave-spin approach}

\author{Daniele Guerci} 
\affiliation{International School for
  Advanced Studies (SISSA), Via Bonomea
  265, I-34136 Trieste, Italy}  

\date{\today} 


\begin{abstract}

We study transport across a magnetic impurity by means of a recently developed slave-spin technique that does not require any constraint. 
 Within a conserving mean-field approximation we find a conductance that displays both the known zero-bias anomaly but also the expected peak at bias of order $U$.  
 We extend the slave-spin mean-field approximation to study the out of equilibrium transient evolution of a quantum dot. We apply the method to investigate the time-evolution of a quantum dot induced by a time-dependent electrochemical potential applied to the contacts. 
Similarly to the time-dependent Gutzwiller approximation, the mean-field slave-spin dynamics is able to capture dissipation in the leads, so that a steady-state is reached after a characteristic relaxation time.

\end{abstract}
\maketitle

\section{Introduction}

Originally observed in magnetic alloys\cite{hewson_1993}, the Kondo effect\cite{Kondo_64,Poor_man}, maybe the simplest collective phenomena due to strong correlations, is now routinely realized in magnetic nanocontacts, either by real magnetic atoms and molecules\cite{Stefan_2014,Stefan_2015,Stefan_2015_PRL} or artificial ones\cite{Kastner_PT,Ashoori_Nat}, e.g. quantum dots, and reveals itself by the so-called zero-bias anomaly\cite{GoldhaberGordon_prl98,Kouwenhoven_science98,Glazman88,NgLee88}.    
It arises by the coupling between a single magnetic atom, such as cobalt, and the conduction electrons of an otherwise non-magnetic metal. Such an impurity typically behaves like a local moment that, due to spin exchange, forms a many-body spin singlet state with the itinerant electrons. 

Unlike magnetic alloys, nanoscale Kondo systems can be driven out of equilibrium by applying charge or spin bias voltages across the devices\cite{Kobayashi_10}. In such a nonequilibrium situation, the interplay between the time dynamics and strong correlation effects makes the theoretical description extremely challenging. To address this problem many innovative approaches has been developed, such as time-dependent numerical renormalization group\cite{tnrg_Anders,Anders_prb06,Andersprl2008}, real time Monte Carlo\cite{Werner_Keldysh_09,Marco_2009}, time-dependent density-matrix renormalization group\cite{Whiteprl2004,Schmitteckertprb2004,Saleur08},  flow equation methods\cite{Kehrein_Kondo,Kehrein_magneticfield,Tomaras_11}, perturbative renormalization group\cite{Pustilnik_1999,Glazman00,Rosch_magneticfield,Metzner_2012,Schoeller2009}, time-dependent variational approaches\cite{Cirac_QI,Nicola_Hugo}, slave-particle techniques\cite{CitroSB,Florencia_Qd,Dong_qd,RAIMONDI_SB} and exact approaches\cite{Andrei06,Bolech2016}. Despite the rich variety of methods, they often become numerically costly at long times, which limit their application to the short times evolution of simple models. However, some of them\cite{Nicola_Hugo,CitroSB}, even if less accurate, are semianalytical methods able to study the full out of equilibrium evolution of realistic systems.

 
To the latter class of approaches belongs the nonequilibrium slave-spin technique for magnetic impurities we present in this paper. By means of a recently developed slave-spin technique\cite{Daniele-SS}, we map without any constraint a single­-orbital Anderson impurity model (AIM),  characterized by a particle-hole symmetric hybridization with the contacts, onto a resonant level model coupled to a single quantum pseudospin. In this suitable representation, a simple self-consistent Hartree-Fock calculation is able to reproduce qualitatively the differential conductance of a single-orbital magnetic impurity both in the small and large bias regimes. Moreover, the slave-spin technique allows to study the full time evolution of magnetic impurities coupled with metallic leads under a nonequilibrium protocol.

The plan of the paper is as follows: we first introduce the AIM to describe a single-orbital magnetic impurity coupled with metallic contacts in section \ref{the_model}. We then present in section \ref{slave-spin_representation} our slave-spin mapping, which allows to compute time-dependent average values without any constraint, details are given in section \ref{fate_cons}.
 In section \ref{td_SS} we present the mean-field approximation for the out of equilibrium dynamics of a single-orbital magnetic impurity. Then, by assuming that the system relaxes after an initial transient, we present, in section \ref{MF_NONeq}, the mean-field approximation for the nonequilibrium steady-state regime. To highlight the importance of the approach presented in this work, section \ref{application} is devoted to the application of the method to transport in magnetic impurities coupled with metallic contacts. In particular, in section \ref{example_1}, we consider the nonequilibrium steady-state induced by applying a constant voltage to the contacts. Furthermore, in section \ref{example_2} we compute within a self-consistent approximation scheme the steady-state differential conductance.
 Finally, section \ref{quasistatic_dyn} is devoted to the analysis of the out of equilibrium evolution induced by a time-dependent voltage applied to the metallic contacts. Technical points of the calculations are given in appendices \ref{K_RLM}, \ref{RPA_eqs} and \ref{K_RLM_transient} at the end of the paper.

\section{The model}
\label{the_model}
We model a single-orbital magnetic impurity coupled to left ($L$) and right ($R$) contacts in terms of an AIM
\be
\begin{split}
H(t;U,V_{g},h)=&H_{dot}(t;U,V_{g},h)+H_{c}+T(t)
\end{split}
\label{AIM}
\ee
where the first term corresponds to an interacting impurity
\be
\label{isolated_dot}
\begin{split}
H_{dot}(t;U,V_{g},h)=&-U\Omega/4-V_{g}(t)(n-1)\\
&-h(t)(n_{\uparrow}-n_{\downarrow}),\\
\end{split}
\ee
where $d_{\sigma}$ is the annihilation operator of an electron state on the impurity, $n_{\sigma}=d^{\dagger}_{\sigma}d_{\sigma}$ the corresponding density, $\Omega=-(2n_{\uparrow}-1)(2n_{\downarrow}-1)$ and $n=n_{\uparrow}+n_{\downarrow}$. In Hamiltonian (\ref{isolated_dot}) $U$ denotes the charging energy, $V_{g}$ the gate potential and $h$ the Zeeman field applied on the dot.
 The non-interacting leads are represented by a free electron gas with half-bandwidth $D$
\be
\label{FEG}
H_{c}=\sum_{a=L,R}\sum_{k\sigma}(\epsilon_{k}-\phi_{a}) c^{\dagger}_{ak\sigma}c_{ak\sigma}, 
\ee
where $\phi_{a}$ is the elettrochemical potential that fixes the number of electrons in each contact, $\phi_{L}=-\phi_{R}$.

Finally, the tunneling coupling between the leads and the central region  is represented by: 
\be
T(t)=\sum_{a=L,R}\sum_{k\sigma} \left(v_{ak}(t)c^{\dagger}_{ak\sigma}d_{\sigma}+H.c.\right)/\sqrt{V},
\label{Op_Hybridisation}
\ee
where $v_{ak}(t)$ is a time-dependent tunneling amplitude, and $V$ is the number of $k$ states. In this article we limit the analysis to the symmetric case where $v_{Lk}(t)=v_{Rk}(t)$.    
 Furthermore, we assume a particle-hole symmetric bath, i.e. for any $\epsilon_{k}$ there exist a $k^{*}$ such that $\epsilon_{k^{*}}=-\epsilon_{k}$ and:
\beS
 \Gamma(-\epsilon,t)= \Gamma(\epsilon,t),
\eeS 
where 
\be
\label{hybr_amplitude}
\Gamma(\epsilon,t)=\pi\sum_{k}|v_{k}(t)|^{2}\delta(\epsilon-\epsilon_{k})/V.
\ee
 
 Under a spin-$\sigma$ particle-hole transformation $\mathcal{C}_{\sigma}$
\be
\Bigg(d_{\sigma}\to d^{\dagger}_{\sigma}\cup\prod_{k}\Big(c_{Lk\sigma}\to -c^{\dagger}_{Rk^{*}\sigma}\cup c_{Rk\sigma}\to -c^{\dagger}_{Lk^{*}\sigma} \Big)\Bigg),
\label{P-H}
\ee
the Hamiltonian (\ref{AIM}) parameters change as follows, 
\be
\label{PH_Transformation}
U\to-U,\quad V_{g}\to\mp h,\quad h\to\mp V_{g},
\ee
where upper and lower signs refer to the action of $\mathcal{C}_{\uparrow}$ and $\mathcal{C}_{\downarrow}$, respectively. The particle-hole transformation (\ref{P-H}) has been defined by mixing $R$ and $L$ contacts to leave the electrochemical potential (\ref{FEG}) invariant.   
 
To study transport across the impurity is convenient to perform the Glazman-Raikh rotation \cite{Glazman88}:
\be
\label{G-R_rotation}
\left(\begin{array}{c}c_{1k\sigma} \\c_{2k\sigma}\end{array}\right)=\frac{1}{\sqrt{2}}\left(\begin{array}{cc}1 & 1 \\1 & -1\end{array}\right)\left(\begin{array}{c}c_{Lk\sigma} \\c_{Rk\sigma}\end{array}\right).
\ee
We notice that the anti-symmetric combination of the electron states in the leads $c_{2k\sigma}$ is fully decoupled from the impurity, while the symmetric combination $c_{1k\sigma}$ remains coupled to $d_{\sigma}$, see Eq. (\ref{Op_Hybridisation}). Thus, the Kondo screening involves only the $c_{1k\sigma}$ variables.
 On the other hand, the current operator is expressed in terms of $c_{2k\sigma}$ only: 
\be
I(t)=-i\sum_{\sigma}\sum_{k}\left(v_{k}(t)c^{\dagger}_{2k\sigma}d_{\sigma}-H.c.\right)/\sqrt{2V},
\label{current_operator}
\ee
where the current operator, defined as $I=(I_{L}-I_{R})/2$ and $I_{a}=\dot{N}_{a}$, is invariant under the particle-hole transformation (\ref{P-H}). 

\subsection{The slave-spin representation}
\label{slave-spin_representation}
In the local magnetic regime, when  $U$ is by far the largest energy scale, charge fluctuations are well-separated in energy from spin ones. However, Hamiltonian (\ref{AIM}) lacks a clear separation between charge and spin degrees of freedom that is desirable in the magnetic moment regime.
To disentangle low and high energy sectors we enlarge the original Hilbert space $\mathscr{H}$ by adding a single quantum pseudospin variable $\sigma$:
\beS
|n\rangle\to|n\rangle\otimes|s\rangle.
\eeS
where $|n\rangle=\{|0\rangle,|\uparrow\rangle,|\downarrow\rangle,|\uparrow\downarrow\rangle\}$ and $|s\rangle=\{|+\rangle,|-\rangle\}$.  Therefore, we encode valence fluctuations, measured by the operator: 
\beS
\Omega = -(2n_{\uparrow}-1)(2n_{\downarrow}-1)=\begin{cases}
      &-1\text{ if \{$|\uparrow\downarrow\rangle$, $|0\rangle$\}},\\
      &+1\text{ if \{$|\uparrow\rangle$, $|\downarrow\rangle$\}},
\end{cases}
\eeS     
in $\sigma^{z}$ by imposing the local constraint that filters the physical subspace out from the enlarged Hilbert space $\mathscr{H}^{*}$:
\beS
 \langle s|\otimes\langle n|\Big(\sigma^{z}\Omega\Big)|n\rangle \otimes|s\rangle=1.
\eeS
Consequently, the eigenstates of $\sigma^{z}$ refer to the presence $\left(|+\rangle\right)$ or the absence $\left(|-\rangle\right)$ of a local magnetic moment in the impurity site. In addition, we introduce two auxiliary fermionic operators $f_{\sigma}$ that annihilate a pseudofermion state on the impurity. The precise relation between the original electrons and the auxiliary degrees of freedom is given by: 
\be
d_{\sigma}=\sigma^{x} f_{\sigma}
\label{mapping}
\ee
ensuring the anticommutation relations $\{d_{\sigma},d^{\dagger}_{\sigma^{\prime}}\}=\delta_{\sigma\sigma^{\prime}}$. 
In the physical subspace, which is selected by the projector 
\be
\label{projector}
\mathbb{P}=\frac{1+\sigma^{z}\Omega}{2},
\ee
 the original model (\ref{AIM}) is equivalent to: 
\be
\begin{split}
H^{*}(t;U,Vg,h)=&H_{c}+\sigma^{x}T(t)+H^{*}_{dot}(t;U,V_{g},h),
\label{RL+S}
\end{split}
\ee
where $H_{c}$ remains unalterated, $T(t)$ is obtained by replacing $d_{\sigma}$ with $f_{\sigma}$ in Eq. (\ref{Op_Hybridisation}), while the dot Hamiltonian is:
\be
\begin{split}
H^{*}_{dot}(t;U,V_{g},h)=&-\frac{U}{4}\sigma^z-V_{g}(t)(1-\sigma^{z})\left(n_{\uparrow}-\frac{1}{2}\right)\\
&-h(t)(1+\sigma^{z})\left(n_{\uparrow}-\frac{1}{2}\right).
\end{split}
\label{H_dotSS}
\ee  
Thus, the original Anderson impurity model is mapped into a resonant level model coupled to the pseudospin $\sigma^{x}$ operator in the presence of a transverse field along the $\sigma^{z}$ component.  
We observe that the Hamiltonian $H^{*}$ possesses a local $Z_{2}$ gauge symmetry generated by the parity transformation $\sigma^{z}\Omega=2\mathbb{P}-1$. Therefore, the quantum dynamics, induced by the operator $\sigma^xT(t)$, couples the singly occupied impurity configuration $\{|\downarrow\rangle,|\uparrow\rangle\}\otimes|+\rangle$  with $\{|0\rangle,|\uparrow\downarrow\rangle\}\otimes|-\rangle$ and  does not mix physical and unphysical subspaces.
 
Finally, we notice that in the physical subspace the current operator reads: 
\be
\label{current_operator_S}
I^{*}(t)=\sigma^{x}I(t)
\ee
where $I(t)$, defined in Eq. (\ref{current_operator}), contains $f_{\sigma}$ pseudofermion operators.

Remarkably, the time-dependent evolution of the AIM, Eq. (\ref{AIM}), can be obtained from the auxiliary model in Eq. (\ref{RL+S}) without any constraint on the enlarged Hilbert space. The proof of this equivalence follows the same steps of the equilibrium case, see Ref. \cite{Daniele-SS}.
  However, we consider valuable to show, in the next section, the possibility to remove the constraint in the time-dependent average value of the charge current, defined in Eq. (\ref{current_operator}).
  
\subsection{Fate of the constraint in the dynamics}
\label{fate_cons}
Without losing generality, we assume the model in Eq. (\ref{RL+S}) prepared at time $t=0$ in thermal equilibrium at temperature $T=1/\beta$:
\beS
\rho(U,V_{g},h)=\frac{e^{-\beta H(U,V_{g},h)}}{Z(U,V_{g},h)},
\eeS
where $Z(U,V_{g},h)=\text{Tr}\left(e^{-\beta H(U,V_{g},h)}\right)$ and the impurity is decoupled from the contacts $v_{k}(0)=0$. For $t>0$ we let the system evolve by suddenly changing the coupling bewteen the bridging region and the leads: $v_{k}(t>0)=v_{k}$.  
 We note that the initial distribution may include a chemical potential bias between $L$ and $R$ contacts.
The average current flowing across the dot (\ref{current_operator}) is defined as:   
\beS
\begin{split}
I(t;U,V_{g},h)&=\Tr\Bigg[\rho(U,V_{g},h)U^{\dagger}(t,0;U,V_{g},h)I \\
&U(t,0;U,V_{g},h)\Bigg],
\end{split}
\eeS
where $U$ is the unitary time evolution operator. 
Since the trace is invariant under similarity transformations and $\mathcal{C}^{\dagger}_{\downarrow}I\mathcal{C}_{\downarrow}=I$, Eq. (\ref{PH_Transformation}) implies:  
\beS
I(t;U,V_{g},h)=I(t;-U,h,V_{g}),
\eeS
and 
\be
\label{Av_Current}
I(t;U,V_{g},h)=\frac{I(t;U,V_{g},h)+I(t;-U,h,V_{g})}{2}.
\ee
Within the slave-spin representation the initial equilibrium distribution is described by 
\beS
\rho^{*}(U,V_{g},h)=\frac{e^{-\beta H^{*}(U,V_{g},h)}}{Z(U,V_{g},h)},
\eeS
and the average value of the current reads 
\beS
\begin{split}
I(t;U,V_{g},h)&=\Tr\Bigg[\rho^{*}(U,V_{g},h)(U^{*})^{\dagger}(t,0;U,V_{g},h) \\
&\sigma^{x}IU^{*}(t,0;U,V_{g},h)\mathbb{P}\Bigg],
\end{split}
\eeS
where the trace is on the enlarged Hilbert space, $\mathbb{P}$, defined in Eq. (\ref{projector}), is the projector in the physical subspace and $U^{*}$ is the time evolution operator generated by $H^{*}$.
In the slave-spin representation (\ref{RL+S}) the role of the p-h symmetry transformation $\mathcal{C}_{\downarrow}$ is simply played by $\sigma^{x}$, so
\beS
\begin{split}
I(t;-U,h,V_{g})&=\Tr\Bigg[\rho^{*}(-U,h,V_{g})(U^{*})^{\dagger}(t,0;-U,h,V_{g}) \\
&\sigma^{x}IU^{*}(t,0;-U,h,V_{g})\mathbb{P}\Bigg]\\
&=\Tr\Bigg[\rho^{*}(U,V_{g},h)(U^{*})^{\dagger}(t,0;U,V_{g},h)\\
&\sigma^{x}IU^{*}(t,0;U,V_{g},h)\sigma^{x}\mathbb{P}\sigma^{x}\Bigg].
\end{split}
\eeS  
Eq. (\ref{Av_Current}) implies:
\beS
\begin{split}
2I(t;U,V_{g},h)&=\Tr\Bigg[\rho^{*}(U,V_{g},h)(U^{*})^{\dagger}(t,0;U,V_{g},h)\\
&\sigma^{x}IU^{*}(t,0;U,V_{g},h)\mathbb{P}\Bigg]\\
&+\Tr\Bigg[\rho^{*}(U,V_{g},h)(U^{*})^{\dagger}(t,0;U,V_{g},h) \\
&\sigma^{x}IU^{*}(t,0;U,V_{g},h)\sigma^{x}\mathbb{P}\sigma^{x}\Bigg].
\end{split}
\eeS
Since $1=\mathbb{P}+\sigma^{x}\mathbb{P}\sigma^{x}$, it readily follows that:
\be
\label{F_SS_current}
\begin{split}
I(t;U,V_{g},h)&=\Tr\Bigg[\frac{e^{-\beta H^{*}(U,V_{g},h)}}{Z^{*}(U,V_{g},h)}(U^{*})^{\dagger}(t,0;U,V_{g},h)\\
&\sigma^{x}IU^{*}(t,0;U,V_{g},h)\Bigg],
\end{split}
\ee
where we have used the equivalence $Z^{*}(U,V_{g},h)=2Z(U,V_{g},h)$. Eq. (\ref{F_SS_current}) states that the time-dependent average value of the current flowing across the impurity (\ref{AIM}) can be computed in the slave-spin representation (\ref{RL+S}) without any constraint.

Following the same line of reasoning, previous result extends to any time-dependent average of physical observables and holds for any nonequilibrium protocol. Thus, we conclude that the out of equilibrium evolution of the original model (\ref{AIM}) can be obtained within the slave-spin representation (\ref{RL+S}) without projecting out unphysical configurations introduced by the mapping (\ref{mapping}).

\section{Time-dependent Mean-field equations}
\label{td_SS}

In this section we present the mean-field approximation to describe the out of equilibrium evolution of a driven magnetic impurity.
The dynamics of the AIM (\ref{AIM}) is governed by the time-dependent Schr\"odinger equation: 
\be
i\partial_{t}|\Psi(t)\rangle=H^{*}(t;U,V_{g},h)|\Psi(t)\rangle,
\label{tdS_exact}
\ee 
where at $t=0$ the system is prepared in the ground state configuration $|\Psi(0)\rangle$ of the initial Hamiltonian Eq. (\ref{RL+S}). 

The mean-field approach consists in approximating\cite{Daniele-SS} the time-dependent wave function $|\Psi(t)\rangle$ with a factorized one product of a fermionic part $|\Phi(t)\rangle$ times a spin one $|\chi(t)\rangle$: 
\be
|\Psi(t)\rangle=|\chi(t)\rangle\otimes|\Phi(t)\rangle.
\label{MF_variationalWF}
\ee 

We notice that the previous approximation is appropriate in the local moment regime, i.e. $U/\Gamma\gg1$, where the two subsystems are characterized by well-separated energy scales. This is indeed the regime we consider hereafter.


 The dynamics of the interacting model (\ref{tdS_exact}) is, thus, reduced to the evolution of a spin degree of freedom:
\be
\label{Ising_dyn}
      \partial_{t}\langle\sigma^{i}(t)\rangle=-2\epsilon_{ijk}\mathcal{B}_{j}(t)\langle\sigma^{k}(t)\rangle, 
\ee 
under a self-consistent time-dependent magnetic field:
\beS
\vec{\mathcal{B}}(t)=\left(-\langle T(t)\rangle,0,\frac{U}{4}+(h(t)-V_{g}(t))\left(\langle n_{\uparrow}(t)\rangle-\frac{1}{2}\right)\right).
\eeS 
Eq.(\ref{Ising_dyn}) is coupled with the Schr\"odinger Eq. for the Slater determinant $|\Phi(t)\rangle$:
\be
\label{Slater_ev}
i\partial_{t}|\Phi(t)\rangle=H^{*}_{f}(t)|\Phi(t)\rangle,
\ee
where the effective fermionic Hamiltonian is    
\be
\label{RL_dyn}
H^{*}_{f}(t)=H_{leads}+\langle\sigma^{x}(t)\rangle T(t)-\lambda_{\uparrow}(t)n_{\uparrow},
\ee
and $\lambda_{\uparrow}(t)=V_{g}(t)(1-\langle\sigma^{z}(t)\rangle)+h(t)(1+\langle\sigma^{z}(t)\rangle)$.
For a given initial configuration, $|\Psi(0)\rangle=|\chi(0)\rangle\otimes|\Phi(0)\rangle$, Eqs. (\ref{Ising_dyn}) and (\ref{Slater_ev}) allow to study the dynamics of the original correlated model in terms of the evolution of a spin $1/2$ coupled with a time-dependent Resonant level model. 

As observed in section \ref{fate_cons}, we emphasize that the nonequilibrium evolution of the Hamiltonian  (\ref{AIM}) can be obtained by the slave-spin representation without any need of local constraints that project out unphsyical configurations introduced by the mapping (\ref{mapping}). The advantages, respect to other slave-particles approaches\cite{CitroSB,RAIMONDI_SB}, are twofold. On one side, we reduce the number of dynamical equations. On the other side, we avoid the mean-field mixing of unphysical and physical subspaces. 

The dynamical Eqs. (\ref{Ising_dyn}) and (\ref{Slater_ev}) are equivalent to the ones obtained by applying the time-dependent Gutzwiller approximation (t-GA) \cite{Marco-PRL} to the AIM \cite{Nicola_Hugo}.
In this regard, the evolution of the time-dependent Gutzwiller parameters resemble the dynamics of the spin variable, while the bath $c_{ak\sigma}$ and the pseudofermion $f_\sigma$ degrees of freedom evolve under a time-dependent self-consistent Hamiltonian (\ref{RL_dyn}).   

 For large time, namely after the transient, we assume that, due to the coupling with infinite contacts, the solution of Eqs. (\ref{Ising_dyn}) and (\ref{Slater_ev}) thermalizes to a steady-state. In order to describe the asymptotic regime we develop, in the next section, the nonequilibrium stationary mean-field approach.
 

\section{Mean-field for the nonequilibrium steady-state}
\label{MF_NONeq}

In this section we discuss the mean-field approximation in the nonequilibrium steady-state. 

Without losing generality, we shall assume that at $t=0$ the contacts are disconnected to the dot but in the presence of a finite bias, so that their distribution functions read: 
\be
\label{contacts_distrib}
\langle c^{\dagger}_{L(R)k\sigma}c_{L(R)k\sigma}\rangle=f_{L(R)}\left(\epsilon_{k}\right)=f\left(\epsilon_{k}\mp\phi/2\right),
\ee
where $\phi$ is the voltage difference applied to the contacts and $f(\epsilon)$ is the Fermi-Dirac distribution function.
Once the tunneling amplitude (\ref{Op_Hybridisation}) is turned on, a time-dependent current starts to flow across the junction accordingly to Eqs. (\ref{Ising_dyn}) and (\ref{Slater_ev}). For large time, namely after the transient, we assume that the system described by the ground-state $|\Psi(t)\rangle$ reaches a stationary state
\be
\label{steady_state}
|\Psi(t)\rangle\to|\Psi\rangle_{st},
\ee
characterized by a constant current.  We observe that Eq. (\ref{steady_state}) is a justified assumption. Indeed, as presented in section \ref{quasistatic_dyn}, the slave-spin mean-field evolution predicts, for large time, the existence of a steady-state due to the coupling of the dot with infinite contacts.    

Following the same reasoning of section \ref{td_SS}, the stationary mean-field approach consists in approximating\cite{Daniele-SS} the ground-state wave function (\ref{steady_state}) with a factorized one: 
\be
|\Psi\rangle_{st}=|\chi\rangle_{st}\otimes|\Phi\rangle_{st},
\label{MF_variationalWF}
\ee 
where $|\Phi\rangle_{st}$ is the fermionic part and $|\chi\rangle_{st}$ the spin one.
At stationarity, the pseudospin degree of freedom is controlled by the Hamiltonian:
\be
\label{HsMF}
\begin{split}
H^{*}_{\sigma}&=-\frac{U}{4}\sigma^{z}+\big\langle T\big\rangle_{st}\,\sigma^{x}\\
&+\left(V_{g}-h\right)\left\langle n_{\uparrow}-\frac{1}{2}\right\rangle_{st}\sigma^{z},
\end{split}
\ee 
where $\langle\cdots\rangle_{st}=\langle\Phi|\cdots|\Phi\rangle_{st}$ and 
\begin{align}
\big\langle T\big\rangle_{st}&=\sqrt{\frac{2}{V}}\sum_{k\sigma}v_{k}\big\langle f^{\dagger}_{\sigma}c_{1k\sigma}+H.c\big\rangle_{st}, \label{effective_coupling1} \\
\langle n_{\uparrow}\rangle_{st}&=\langle f^{\dagger}_{\uparrow}f_{\uparrow}\rangle_{st}, \label{effective_coupling2} 
\end{align}
are expectation values in the fermionic steady-state wave function.
The ground-state of (\ref{HsMF}) is identified by:
\be
\label{spin_gs}
\begin{split}
\langle\sigma^{x}\rangle_{st}&\equiv\sin\theta=\frac{\mathcal{B}_{x}/\mathcal{B}_{z}}{\sqrt{1+\left(\mathcal{B}_{x}/\mathcal{B}_{z}\right)^{2}}},\\
\langle\sigma^{z}\rangle_{st}&\equiv\cos\theta=\frac{1}{\sqrt{1+\left(\mathcal{B}_{x}/\mathcal{B}_{z}\right)^{2}}},
\end{split}
\ee
where for convenience we have introduced the self-consistent magnetic field:
\be
\mathcal{B}=\left(-\big\langle T\big\rangle_{st},0,\frac{U}{4}-\left(V_{g}-h\right)\left\langle n_{\uparrow}-\frac{1}{2}\right\rangle_{st}\right).
\ee
The fermionic problem is, thus, reduced to find the steady-state ground-state of the quantum Hamiltonian
 \be
 \label{mf_F_neq}
 \begin{split}
 H^{*}_{f}&=H_{c}+\sin\theta \sum_{k\sigma}\sqrt{\frac{2}{V}}v_{k}\left(c^{\dagger}_{1k\sigma}f_{\sigma}+H.c.\right)\\
 &-\lambda_{\uparrow}n_{\uparrow}
 \end{split}
 \ee
 where $c_{1k\sigma}$ is introduced in the aforementioned unitary transformation (\ref{G-R_rotation}) and 
 \beS
 \label{lambda_up}
 \lambda_{\uparrow}=h(1+\cos\theta)+V_{g}(1-\cos\theta).
 \eeS

Since we deal with a nonequilibrium situation we work in the framework of the Keldysh technique, as employed in the literature \cite{Rammer_book_noneq,HAUG_JAUHO,Arseev_2015}. Eq. (\ref{effective_coupling1}) requires the evaluation of the lesser Green's function $G^{<}_{1kf\sigma}(t,t)=i\langle  f^{\dagger}_{\sigma}(t)c_{1k\sigma}(t)\rangle$, which, by means of the Dyson's Eq., can be expressed in terms of the dressed Green's function of the $f_{\sigma}$ pseudofermions and the free Green's function of the contacts. Instead, Eq. (\ref{effective_coupling2}) can be expressed in terms of the pseudofermions Green's function only.
By performing straightforward calculations, that are summarized in appendix \ref{K_RLM}, we obtain:
\begin{align}
\langle T\rangle_{st}&=\frac{2}{\sin\theta}\sum_{\sigma}\int d\epsilon (\epsilon+\lambda_{\sigma})f_{\text{neq}}(\epsilon) A_{f\sigma}(\epsilon), \label{effective_couplings_1T}\\
\langle n_{\uparrow}\rangle_{st}&=\int d\epsilon f_{\text{neq}}(\epsilon) A_{f\uparrow}(\epsilon),\label{effective_couplings_1n}
\end{align}
where the nonequilibrium distribution on the impurity is $f_{\text{neq}}(\epsilon)=(f_{L}(\epsilon)+f_{R}(\epsilon))/2$ and the $f_{\sigma}$ pseudofermion spectral function reads 
\beS
\label{A_f}
A_{f\sigma}(\epsilon)=\frac{1}{\pi}\frac{-\text{Im}\Sigma^{R}_{f\sigma}(\epsilon)}{(\epsilon+\lambda_{\sigma}-\text{Re}\Sigma^{R}_{f\sigma}(\epsilon))+\text{Im}\Sigma^{R}_{f\sigma}(\epsilon)^{2}}.
\eeS
Within the mean-field approximation, the $f_{\sigma}$ pseudofermions self-energy is given by:
\beS
\label{f_self_e}
\Sigma^{R}_{f\sigma}(\omega)=2\sin^{2}\theta\int\frac{ d\epsilon}{\pi}\frac{\Gamma(\epsilon)}{\omega-\epsilon+i0^{+}}, 
\eeS
where the factor of 2 counts the presence of two different leads, while the hybridization function $\Gamma(\epsilon)$ is defined in Eq. (\ref{hybr_amplitude}).

Given the spectral properties of the contacts, i.e. $\Gamma(\epsilon)$, Eqs. (\ref{effective_couplings_1T}) and (\ref{effective_couplings_1n})  give an analytic expressions for the  effective magnetic field $\mathcal{B}$, which depends on the steady-state average $\langle \sigma^{x}\rangle_{st}$. Therefore, we close the set of mean-field equations and the steady-state variational ground-state is obtained by solving: 
\be
\label{mf_SCE}
\sin\theta=\frac{\mathcal{B}_{x}(\theta)/\mathcal{B}_{z}(\theta)}{\sqrt{1+\left(\mathcal{B}_{x}(\theta)/\mathcal{B}_{z}(\theta)\right)^{2}}}
\ee
that corresponds to a root-finding problem $g(\theta)=0$ in a single angular variable $\theta$.

Before concluding the section, we observe that the nonequilibrium steady-state self-consistent Eq. (\ref{mf_SCE}) is equivalent to the one obtained with the out of equilibrium Gutzwiller approach for quantum dots\cite{Nicola1}. 
However, in comparison with the latter approach, the slave-spin method has the advantage of allowing one to use the machinery of quantum field theory, i.e. Wick's theorem, to improve mean-field results by including fluctuations.

\section{Application to transport thorugh a magnetic impurity}
\label{application}

The last section of this work is devoted to the application of the method, developed in sections \ref{td_SS} and \ref{MF_NONeq}, to study the nonequilibrium dynamics of a magnetic impurity coupled with metallic contacts. To highlight the importance of our formulation here we consider the simple case $V_{g}=h=0$, and we take the wide-band limit (WBL).  Moreover, we will firstly analyze the steady-state regime by computing the nonequilibrium ground-state and the differential conductance as a function of the voltage applied to the contacts. Then, we will study the out of equilibrium evolution induced by a slowly varying time-dependent voltage. 

\subsection{The steady-state solution in the wide band limit}
\label{example_1}

Initially, we assume the dot disconnected by the leads, which are prepared at two different chemical potential $\pm\phi/2$, so that their initial distribution function is described by Eq. (\ref{contacts_distrib}). Once the tunneling amplitude is turned on, after the initial transient, the steady-state Hamiltonian, that describes the quantum pseudospin degree of freedom, is given by: 
\beS
H^{*}_{\sigma}=-\frac{U}{4}\sigma^{z}+\langle T\rangle_{st}\sigma^{x}.
\eeS
In the wide-band limit, where $\Gamma(\epsilon)=\Gamma_{0}$, the $f$ electron self-energy reduces to 
\be
\label{f_self_e_WBL}
\Sigma^{R}_{f\sigma}(\omega)=-i2\Gamma_{0}\sin^{2}\theta,
\ee
and we readily find that
\be
\langle T\rangle_{st}=-\frac{4\Gamma}{\pi\sin\theta}\log\frac{D}{\sqrt{\Gamma^{2}+\phi^{2}/4}},
\ee 
where $\Gamma$ is the renormalized hybridization amplitude $\Gamma=2\Gamma_{0}\sin^{2}\theta$.
The steady-state variational ground-state is obtained by solving the self-consistent equation: 
\be
\label{mf_SCE_1}
\sin\theta=-\frac{4\langle T\rangle_{st}/U}{\sqrt{1+\left(4\langle T\rangle_{st}/U\right)^{2}}}.
\ee

 For large $U$, and $\phi\ll\Gamma$, the solution of the self-consistent Eq. (\ref{mf_SCE_1}) for $\Gamma$ reads:
\be
\label{TK_phi}
\Gamma(\phi)\simeq\Gamma(0)-\frac{\phi^{2}}{8\Gamma(0)},  
\ee
where 
\beS
\label{TK_0}
\Gamma(0)=D\exp\left[-\frac{\pi U}{16(2\Gamma_{0})}\right]
\eeS
is the same as in slave-boson mean-field theory, and can be associated with the Kondo temperature $T_{K}$, though overestimated respect its actual value \cite{Pierpaolo}.
As shown in Eq. (\ref{TK_phi}) the effect of an external voltage $\phi$, within mean-field approximation, is to reduce the equilibrium value of the renormalized hybridization $\Gamma(0)$. Moreover, the mean-field steady-state breaks spontaneously the $Z_{2}$ gauge symmetry by choosing one of the two degenerate minima $\langle\sigma^{x}\rangle_{st}\neq0$, as already observed in the equilibrium case Ref.\cite{Pierpaolo}.

At the steady-state variational minimum we can compute the average value of the current:
\be
\label{AV_current}
\langle I\rangle_{st}=-\frac{i}{\sqrt{2V}}\sum_{k\sigma}v_{k}\left(\langle c^{\dagger}_{2k\sigma}\sigma^{x}f_{\sigma}\rangle_{st}-c.c.\right)
\ee 
that involves the evaluation of the two-particle correlation function $G^{<}_{x\cdot2k\sigma}(t,t^{\prime})=i\langle c^{\dagger}_{2k\sigma}(t^{\prime})\sigma^{x}(t)f_{\sigma}(t)\rangle_{st}$.  
In a consistent approximation scheme the self-energy corrections have to be included in two-particle correlation functions through the Bethe-Salpeter equation. 
In the next section, by means of the Abrikosov representation\cite{Abrikosov_Fermions} of the pseudospin variable $\sigma$, we readily compute the average value of the current (\ref{AV_current}) consistently with the mean-field approximation (\ref{MF_variationalWF}).

\subsection{The steady-state current within a self-consistent mean-field approximation}
\label{example_2}
 
To perform a self-consistent calculation of the current, Eq. (\ref{AV_current}), we introduce a couple of fermionic operators $\psi$ corresponding to the pseudospin operator $\vec{\sigma}$ according to the formula\cite{Abrikosov_Fermions}:
\be
\label{Abrikosov_Sf}
\psi^{\dagger}_{\alpha}\sigma^{i}_{\alpha\beta}\psi_{\beta}=\hat{\sigma}^{i}
\ee
where the upper index $i=1,2,3$ denotes the Pauli matrices, while $\alpha,\beta=\pm$. 
The fermion substitution Eq. (\ref{Abrikosov_Sf}) introduces two additional configurations $(0,0)$ and $(1,1)$ to the two dimensional Hilbert space of the $\sigma$-matrices, which is composed by $(1,0)$ and $(0,1)$. However, in the case of spin $S=1/2$ the unphysical configurations are automatically excluded since physical quantities involve only averages of products of $\hat{\sigma}^{i}$, which have the property of giving zero when acting on the non-physical states $(0,0)$ or $(1,1)$.
\begin{figure}
\begin{center}
\includegraphics[width=0.48\textwidth]{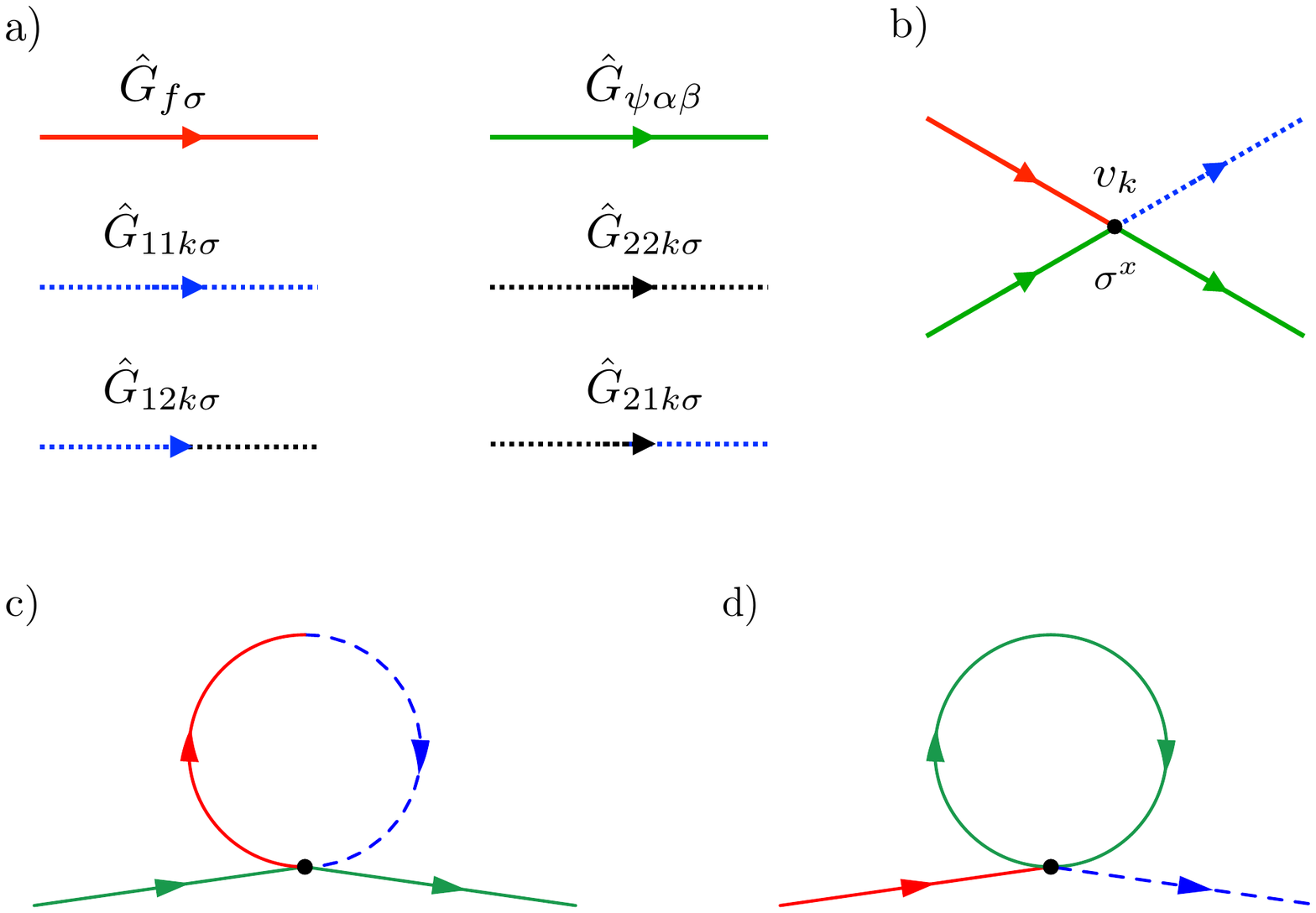}
\caption{a) Bare Green's functions. b) Bare interaction. Hartree-Fock self-energy diagrams corresponding to the slave-spin mean-field approximation: c) elastic scattering between $f_{\sigma}$ and $c_{1k\sigma}$ fermions renormalized by $\langle\psi^{\dagger}_{\alpha}\sigma^{x}_{\alpha\beta}\psi_{\beta}\rangle$, d) $\psi$ fermions self-energy determined by valence fluctuations induced by the hybridization operator $T$.}
\label{Lines}
\end{center}
\end{figure}

\begin{figure}
\begin{center}
\includegraphics[width=0.48\textwidth]{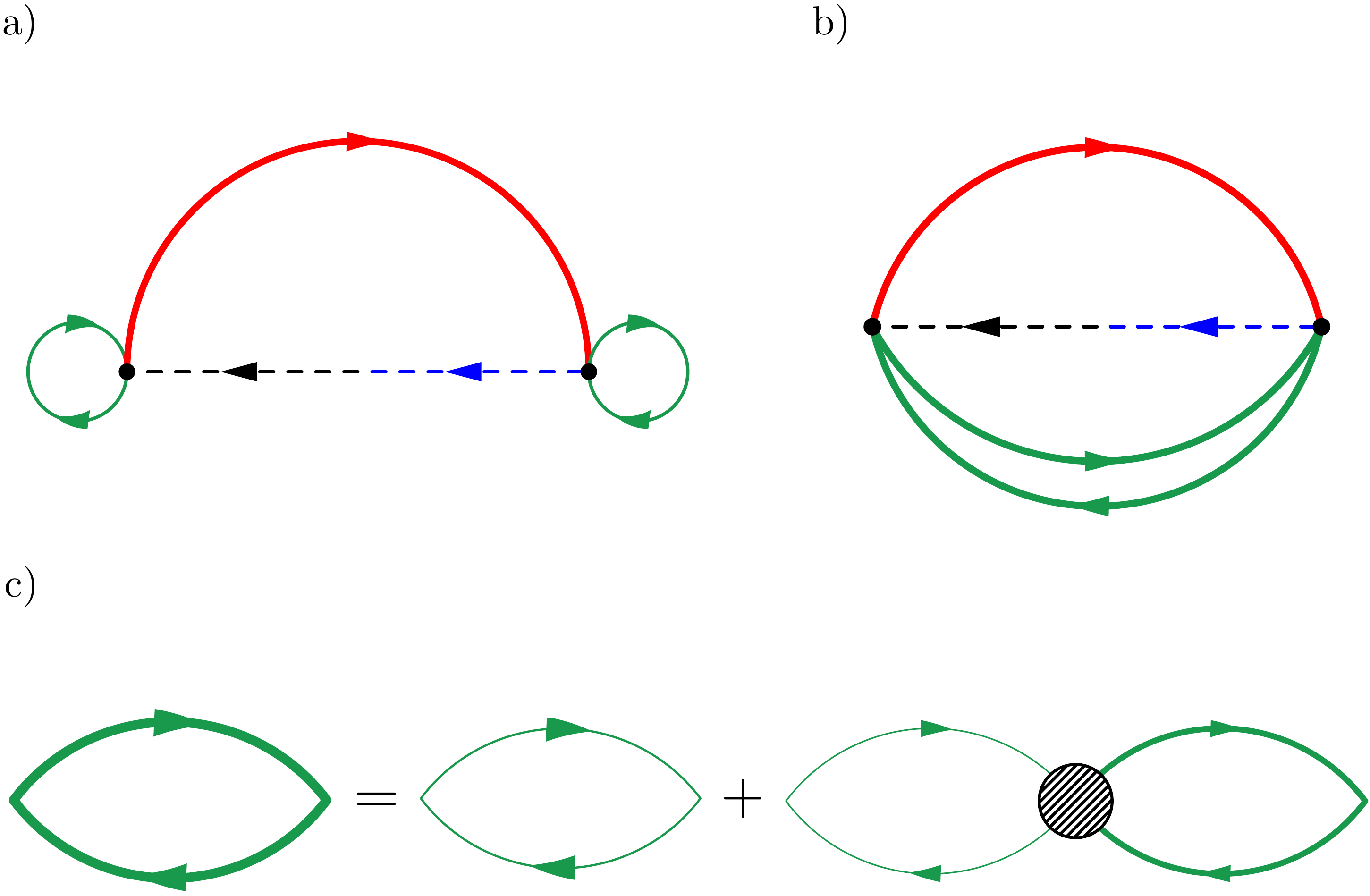}
\vspace{-0.6cm}
\caption{Feynman diagrams contributing to the average value of the current. Top panel: a) $\langle I_{f}\rangle_{st}$ low-energy contribution to the current given by a Resonant level model with renormalized hybridization amplitude, b)  $\langle\delta I\rangle_{st}$ is determined by the convolution of the low-energy fermions with the valence fluctuations described by $\Pi_{xx}$. 
Lower panel: c) Dyson's Eq. for the $\Pi_{xx}$ propagator.}
\label{diagrammi_transport}
\end{center}
\end{figure}

In this representation, the hybridization term in Eq. (\ref{Op_Hybridisation}) becomes the four-leg fermionic interaction vertex depicted in Fig. \ref{Lines} b).    
The Hartree-Fock approximation corresponds to the mean-field decoupling presented in section \ref{MF_NONeq},  and is described by the self-energy diagrams in Figs. \ref{Lines} c) and d).
The average value of the current reads:
\beS
\label{AV_current_1}
\langle I\rangle_{st}=-\frac{i}{\sqrt{2V}}\sum_{k\sigma}v_{k}\left(\langle c^{\dagger}_{2k\sigma}\psi^{\dagger}_{\alpha}\sigma^{x}_{\alpha\beta}\psi_{\beta}f_{\sigma}\rangle_{st}-\text{c.c.}\right)
\eeS 
and implies the evaluation of the two-particle correlation function $\langle c^{\dagger}_{2k\sigma}\psi^{\dagger}_{\alpha}\sigma^{x}_{\alpha\beta}\psi_{\beta}f_{\sigma}\rangle_{st}$.  Therefore, consistently with the slave-spin mean-field decoupling the current is made up of two contributions, Figs. \ref{diagrammi_transport} a) and b):
\be
\label{AV_current_2}
\langle I\rangle_{st}=\langle I_{f}\rangle_{st}+\langle \delta I\rangle_{st}
\ee
where the former, $\langle I_{f}\rangle_{st}$, involves only the low-energy pseudofermion degree of freedom, and can be obtained by straightforward calculations summarized in appendix \ref{K_RLM}. Here, we report the final result in the WBL: 
\be
\label{I_f}
\langle I_{f}\rangle_{st}=
2\Gamma(\phi)\frac{2e}{h}\arctan\left(\frac{e\phi}{2\Gamma(\phi)}\right),
\ee 
where $e$ is the elementary charge and $h$ the Planck's constant.

\begin{figure}
\begin{center}
\includegraphics[width=0.48\textwidth]{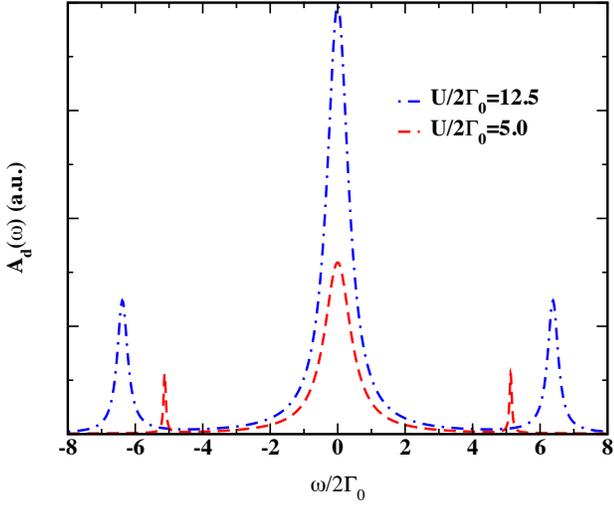}
\vspace{-0.6cm}
\caption{Physical $d_{\sigma}$ electrons spectral function $A_{d}(\omega)$ computed at equilibrium, $\phi=0$, for $U/D=0.1$ and $U/2\Gamma_{0}=12.5, 5.0$. In addition to the low energy Abrikosov-Suhl or Kondo resonance $A_{d}(\omega)$ presents high energy side-bands.}
\label{Ad}
\end{center}
\end{figure}

Instead, the latter term in Eq. (\ref{AV_current_2}) takes into account the contribution of valence fluctuations and can be expressed as 
\be
\label{delta_I}
\langle\delta I\rangle_{st}=-\frac{4\Gamma_{0}e}{h}\int d\omega\left(f_{L}(\omega)-f_{R}(\omega)\right)\text{Re}\,\mathcal{K}(\omega)
\ee 
where the kernel $\mathcal{K}(\omega)$ is given by: 
\beS
\label{Kappa_retarded}
\begin{split}
\mathcal{K}(\omega)&=\int\frac{d\epsilon}{2\pi}\Big[\Pi^{<}_{xx}(\epsilon)G^{R}_{f}(\omega-\epsilon)\\
&+\Pi^{R}_{xx}(\epsilon)G^{R}_{f}(\omega-\epsilon)+\Pi^{R}_{xx}(\epsilon)G^{<}_{f}(\omega-\epsilon)\Big],  
\end{split}
\eeS
where $\Pi_{xx}$ is the $\psi$ fermion spin-correlation function, for more details we refer to appendix \ref{RPA_eqs}. 
Consistently with the Hartree-Fock approximation $\Pi_{xx}$ satisfies the Dyson's Eq. in Fig. \ref{diagrammi_transport} c), whose solution for the retarded component reads:
\be
\label{DysonBOSE_R_Main}
\Pi_{xx}^{R}(\omega)=\frac{1}{\left[{\Pi^{0}}^{R}_{xx}(\omega)\right]^{-1}-\Sigma_{xx}^{R}(\omega)},
\ee
and the lesser component:
\be
\label{DysonBOSE_L_Main}
\Pi^{<}_{xx}(\omega)=\Pi_{xx}^{R}(\omega)\Sigma_{xx}^{<}(\omega)\Pi_{xx}^{A}(\omega),
\ee
where ${\Pi^{0}}^{R}_{xx}(\omega)=2\omega_{0}\cos^{2}\theta/(\omega^{2}-\omega^{2}_{0})$ and $\Pi^{A}_{xx}(\omega)=\left[\Pi^{R}_{xx}(\omega)\right]^{*}$. The self-energies appearing in Eqs. (\ref{DysonBOSE_R_Main}) and (\ref{DysonBOSE_L_Main}) are obtained by contracting the four-leg vertex in Fig. \ref{Lines} b), details can be found in appendix \ref{RPA_eqs}.
Specifically, the self-energy $\Sigma_{xx}(\omega)$ allows to reconstruct incoherent side bands characterized by a width of the order of the bare hybridization $\Gamma_{0}$ and centered around $\pm U/2$ as shown in Fig. \ref{Ad}.

\begin{figure}
\begin{center}
\includegraphics[width=0.48\textwidth]{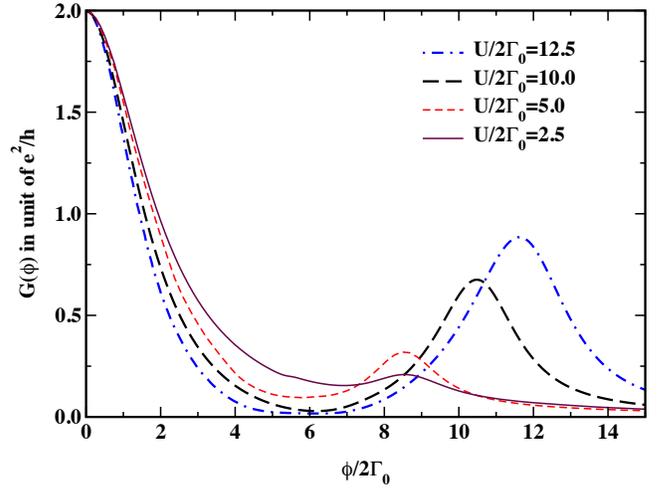}
\caption{Differential conductance as a function of the applied voltage $\phi/2\Gamma_{0}$ for $U/D=0.1$ and different hybridization amplitudes $2\Gamma_{0}$.}
\label{G_voltage}
\end{center}
\end{figure}

 Numerical integration of Eq. (\ref{delta_I}) permits to compute the differential conductance
 \beS
 \label{Diff_conductance}
 G(\phi)=\frac{d\langle I\rangle_{st}}{d\phi},
 \eeS
 which is shown in Fig. \ref{G_voltage}. We observe two distinct contributions: (i)  the well-known zero-bias anomaly which derives from the Kondo peak at the Fermi level and controls the low-bias behavior and (ii) an incoherent one, which mainly contributes to the large bias features of the conductance.
 
To compare our result for $G(\phi)$ with the universal behavior of the conductance in the Kondo regime, obtained with renormalization group approach in Refs.\cite{Glazman04,EranSela}, we expand $\langle I\rangle_{st}$ around $\phi/\Gamma\ll1$ obtaining:
\be
\label{G_low_bias}
G(\phi)=\frac{2e^{2}}{h}\left[1-\frac{1}{4}\left(\frac{\phi}{\Gamma}\right)^{2}\right].
\ee
In agreement with our self-consistent Hartree-Fock approximation, Eq.(\ref{G_low_bias}) reproduces exactly the $\phi^{2}$ contribution given by the phase shift, while neglects the contribution from the residual scattering among low-energy quasiparticles\cite{Nozieres_FL}.  We believe that, in the slave-spin representation, the latter contribution comes from  vertex corrections, that are not included in our perturbative calculation.

\subsection{Adiabatic dynamic induced by a time-dependent voltage}
\label{quasistatic_dyn}

Physically, applying a time-dependent voltage between the source and the drain contacts means that the single-particle energies become time-dependent: $\epsilon_{k}\to\epsilon_{k}-\phi_{a}(t)$ (here $a$ label refers to the left $L$ or right $R$ lead) \cite{Jauho_1994}. Starting, at $t=0$, from an equilibrium configuration characterized by $\phi_{L}=\phi_{R}=0$ ($N_{L}=N_{R}$) and a finite tunneling amplitude $v_{k}$, we consider the evolution induced by a time-dependent electrochemical potential:
\be
\label{electrochemical_pot}
\phi_{L}(t)=\theta(t)\phi\frac{1-e^{-t/t^{*}}}{2},\quad\phi_{R}(t)=-\phi_{L}(t),
\ee
where $t^{*}$ is the characteristic time scale of the external perturbation, $\phi$ is the asymptotic value of the voltage and $\theta(t)$ is the Heaviside step function such that $\phi_{L}(t)=0$ for $t\le0$. Here we consider the WBL  analogously to the steady-state analysis.
The dynamic of the pseudospin variable is:
 \be
\label{Ising_dyn_mu}
\begin{split}
      \partial_{t}\langle\sigma^{x}(t)\rangle&=U\langle\sigma^{y}(t)\rangle/2, \\
            \partial_{t}\langle\sigma^{y}(t)\rangle&=-2\langle T(t)\rangle\langle\sigma^{z}(t)\rangle-U\langle\sigma^{x}(t)\rangle/2,\\
      \partial_{t}\langle\sigma^{z}(t)\rangle&=2\langle T(t)\rangle\langle\sigma^{y}(t)\rangle,\\ 
\end{split}
\ee
where the time-dependent average value of the hybridization is given by:
 \be
 \label{T_2}
\langle T(t)\rangle=\frac{2}{\langle\sigma^{x}(t)\rangle}\text{Im}\left[\int\frac{d\epsilon}{\pi}\Sigma^{<}_{f}(t,\epsilon)\star G^{A}_{f}(t,\epsilon)\right].
 \ee
In this case (\ref{T_2}), the normal product is substituted with $\star=\exp\left[i(\overleftarrow{\partial}_{\epsilon}\overrightarrow{\partial}_{t}-\overleftarrow{\partial}_{t}\overrightarrow{\partial}_{\epsilon})/2\right]$, while $\Sigma^{<}_{f}(t,\epsilon)$ and $G^{A}_{f}(t,\epsilon)$ are the Wigner transform of the lesser component of the self-energy and the advanced Green's function of the $f_{\sigma}$ pseudofermions, for more details we refer to appendix \ref{K_RLM_transient}.

 In the following, we consider an external perturbation $\phi(t)$, which is a slowly varying function of time compared to the characteristic scales of the equilibrium state, i.e. $t^{*}T_{K}\gg1$. Therefore, we can assume that the temporal inhomogeneity is weak and only lowest-order terms in the variation are kept, the so-called gradient expansion \cite{Rammer_book_noneq,HAUG_JAUHO}.
\begin{figure}
\begin{center}
\includegraphics[width=0.48\textwidth]{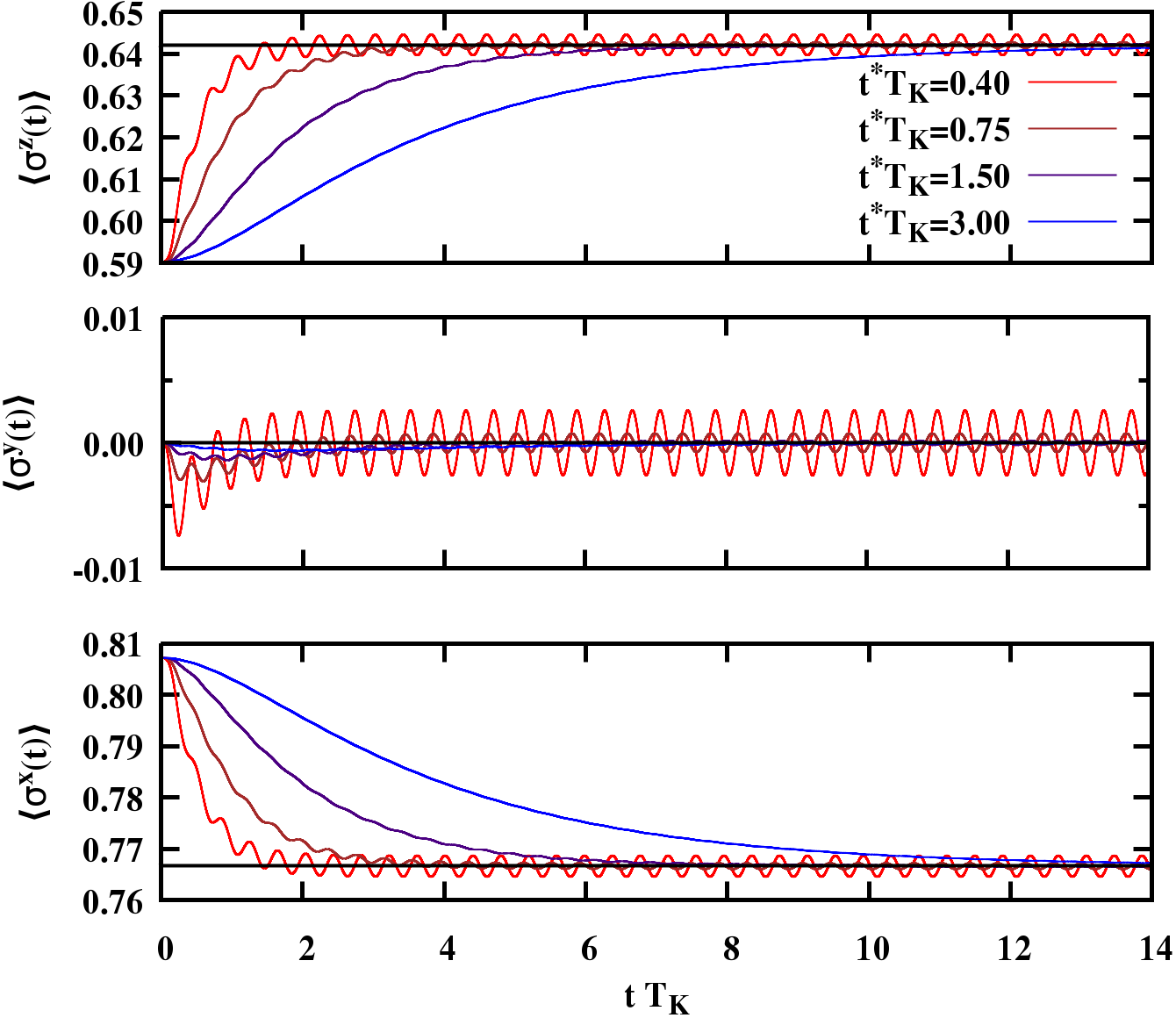}
\vspace{-0.6cm}
\caption{From top to bottom: evolution of  $\langle\sigma^{z}(t)\rangle$, $\langle\sigma^{y}(t)\rangle$ and $\langle\sigma^{x}(t)\rangle$ as a function of $t\,T_{K}$ for several values of the external voltage time scale $t^{*}$, $U/D=0.1$, $2\Gamma_{0}/U=0.06$ and $\phi/U=0.05$. Solid black line represents the steady-state result for the same set of parameters.}
\label{adiabatic_dynamics}
\end{center}
\end{figure}
\begin{figure}
\begin{center}
\includegraphics[width=0.48\textwidth]{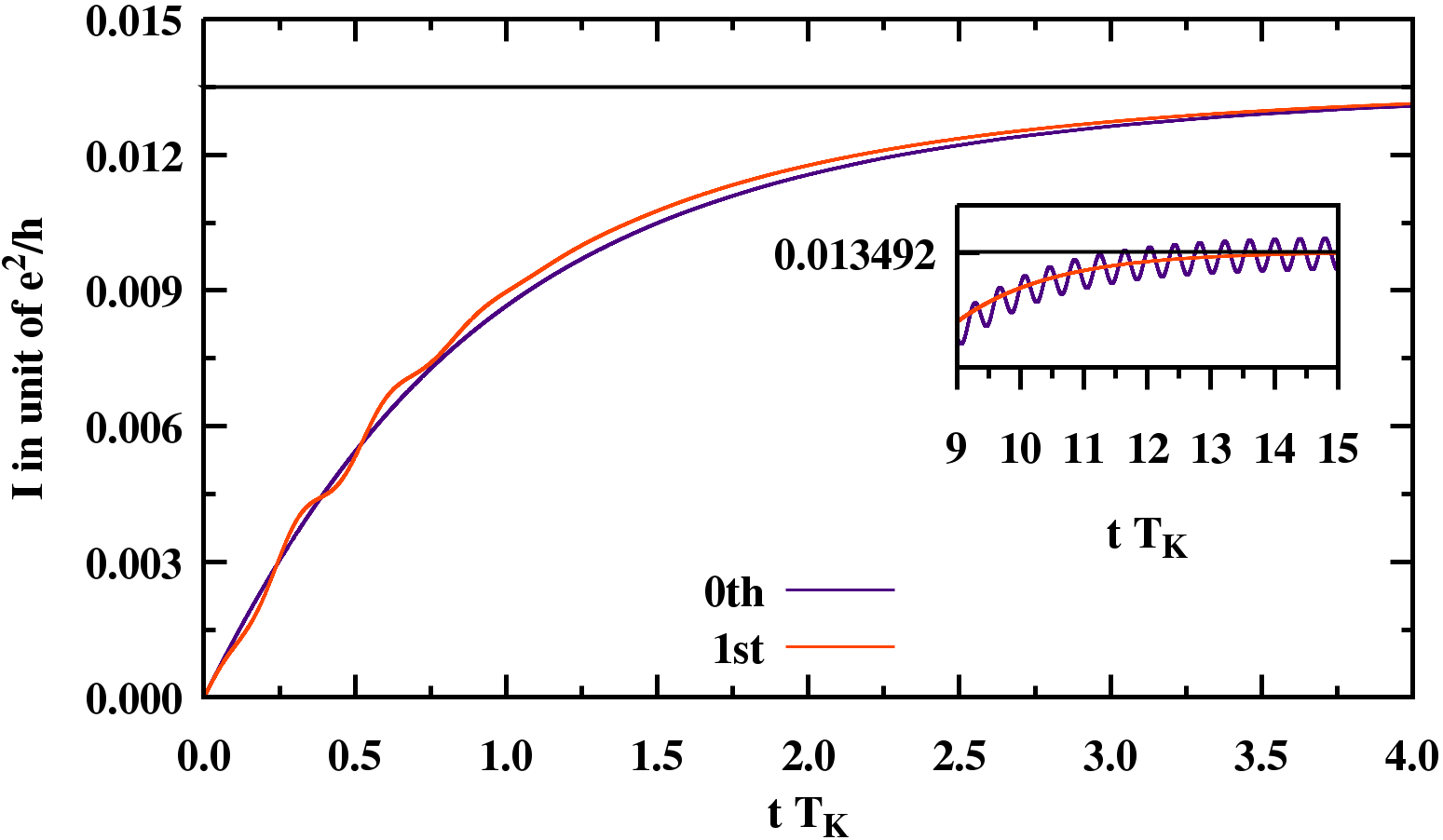}
\vspace{-0.6cm}
\caption{Time-dependent average value of the current as a function of $t\,T_{K}$ for  $t^{*}T_{K}=1.5$, $U/D=0.1$, $2\Gamma_{0}/U=0.06$ and $\phi/U=0.05$. Orange and purple lines represent the evolution of the current obtained within first and zeroth order in the gradient expansion. As shown from the inset, first order corrections to the quasistatic approximation introduce relaxation processes that suppress the residual oscillations.}
\label{Ic_dynamics}
\end{center}
\end{figure}
To the first-order in the temporal variation we have: 
\be
\label{gradient_exp}
\begin{split}
\langle T(t)\rangle&\simeq\frac{2}{\langle\sigma^{x}(t)\rangle}\text{Im}\int\frac{d\epsilon}{\pi}\Bigg[\Sigma^{<}_{f}(t,\epsilon) G^{A}_{f}(t,\epsilon)\\
&+\frac{i}{2}\Big\{\Sigma^{<}_{f}(t,\epsilon), G^{A}_{f}(t,\epsilon)\Big\}_{\epsilon,t}\Bigg]\\
&=\langle T(t)\rangle^{(0)}+\langle T(t)\rangle^{(1)}
\end{split}
\ee 
where $\{f,g\}_{\epsilon,t}=\partial_{\epsilon}f\partial_{t}g-\partial_{t}f\partial_{\epsilon}g$, more details can be found in Appendix \ref{K_RLM_transient}.
  
The evolution of the pseudospin variable induced within the zeroth order in the gradient expansion Eq. (\ref{gradient_exp}) is displayed in Fig. \ref{adiabatic_dynamics}.  In the limit of $t^{*}T_{K}\gg1$ we observe, as expected, the quasistatic dynamic, i.e. the system stays in equilibrium at all times and follows the change of $\mu(t)$ adiabatically. However, for any smaller value of $t^{*}T_{K}$  the dynamics is characterized by persistent oscillations, that become, eventually, centered around the steady-state result represented by the solid black line.  

Remarkably, first-order correction, given by the latter term in Eq. (\ref{gradient_exp}), introduces a relaxation mechanism and the dynamic converges to the expected stationary regime. This is shown in Fig. \ref{Ic_dynamics}, where we compare the time-dependent average value of the current obtained within the zeroth and first order in the gradient expansion.

\section{Conclusions}

We have shown that the out of equilibrium evolution of a single-orbital AIM (\ref{AIM}) can be calculated in the slave-spin representation (\ref{RL+S}) without any constraint on the enlarged Hilbert space. The advantages of the new representation are twofold. On one side, we disentangle charge and spin degrees of freedom. On the other side, we avoid the mean-field mixing of unphysical and physical subspaces, that affects the time evolution of other slave-particle techniques. In the steady-state regime the self-consistent Hartree-Fock decoupling is able to predict properties of the model even deep inside the large-$U$ Kondo regime, specifically, the conductance shows both the known zero-bias anomaly but also the expected peak at bias of order $U$.
Furthermore, we have extended the slave-spin approach to study the transient dynamic of a driven magnetic impurity.
 By means of a time-dependent Hartree-Fock calculation, in the adiabatic regime, we prove that, at first-order in the gradient expansion, the current relaxes to the steady state value after an initial transient.
 
Finally, we mention that the technique we have proposed can be applied to study the out of equilibrium dynamics of multi-orbitals magnetic impurities by using the generalized mapping presented in Ref.\cite{Daniele-SS}.

\section*{Acknowledgments}

I am grateful to Michele Fabrizio for insightful discussions that allowed me to clarify several important points related to this work  and for a careful reading of this manuscript. Furthermore, I thank  Roberto Raimondi, Francesco Grandi, Massimo Capone, Laura Fanfarillo, Valentina Brosco, Maria Florencia Ludovico and Adriano Amaricci for constructive discussions on the manuscript. We acknowledge support from the H2020 Framework Programme under ERC Advanced Grant No. 692670 FIRSTORM.

\appendix
\section{The effective Resonant level model in the steady-state regime}
\label{K_RLM}

In this section we derive analytic expressions for the hybridization Eq. (\ref{effective_coupling1}) and the current Eq. (\ref{I_f}). Morover, we compute the Keldysh's components of the $f_{\sigma}$ and $\psi$ fermion Green's function within Hartree-Fock approximation.  
\paragraph{$f_{\sigma}$ pseudofermion Green's function}
The unperturbed retarded and advanced Green's functions of the contacts are 
\beS
\label{12_GRA}
\begin{split}
G^{R/A}_{11\sigma}(\epsilon,k)&=G^{R/A}_{22\sigma}(\epsilon,k)=\frac{1}{\epsilon-\epsilon_{k}\pm i0^{+}},\\
G^{R/A}_{12\sigma}(\epsilon,k)&=G^{R/A}_{21\sigma}(\epsilon,k)=0, 
\end{split}
\eeS
and 
\beS
\label{12_GRA}
\begin{split}
G^{<}_{11\sigma}(\epsilon,k)&=G^{<}_{22\sigma}(\epsilon,k)=2i\pi\delta(\epsilon-\epsilon_{k})\frac{f_{L}(\epsilon)+f_{R}(\epsilon)}{2},\\
G^{<}_{12\sigma}(\epsilon,k)&=G^{<}_{21\sigma}(\epsilon,k)=2i\pi\delta(\epsilon-\epsilon_{k})\frac{f_{L}(\epsilon)-f_{R}(\epsilon)}{2},
\end{split}
\eeS
where we have already performed the rotation in Eq. (\ref{G-R_rotation}).
In terms of the matrix representation
\be
\label{Matrix_Keldysh}
\hat{G}=\begin{pmatrix}
      G^{R}& G^{<}    \\
      0       & G^{A} 
\end{pmatrix}
\ee
the Dyson's equation for the $f_{\sigma}$ pseudofermion Green's function on the Keldysh's contour is: 
\be
\label{Dyson_fRA}
\hat{G}_{f\sigma}=\hat{G}^{0}_{f\sigma}+\hat{G}^{0}_{f\sigma}\cdot\hat{\Sigma}_{f}\cdot \hat{G}_{f\sigma}
\ee
where $\hat{G}_{f\sigma}$ is the dressed Green's function and $\hat{G}^{0}_{f\sigma}$ the unperturbed one.
In Eq. (\ref{Dyson_fRA}) we use a notation where the product $\cdot$ is interpreted as a matrix product in the internal variables (time and Keldysh's indeces).  In the stationary regime the time translational invariance is restored, thus, by taking the Fourier transform of Eq. (\ref{Dyson_fRA}) we obtain:
 \be
 \label{RA_Dyson}
 G^{R/A}_{f\sigma}(\epsilon)=\frac{1}{\epsilon+\lambda_{\sigma}-\Sigma^{R/A}_{f\sigma}(\epsilon)}
 \ee 
and 
\be
\label{<_Dyson}
 G^{<}_{f\sigma}(\epsilon)=G^{A}_{f\sigma}(\epsilon) \Sigma^{<}_{f\sigma}(\epsilon) G^{R}_{f\sigma}(\epsilon).
\ee
Within mean-field approximation the self-energy of the $\Sigma_{f\sigma}$ reads:
\beS
\label{SE_RA}
\begin{split}
\Sigma^{R/A}_{f\sigma}(\epsilon)&=\langle\sigma^{x}\rangle^{2}_{st}\frac{2}{V}\sum_{k}v^{2}_{k}G^{R/A}_{11\sigma}(\epsilon,k)\\
&=2\langle\sigma^{x}\rangle^2_{st}\int \frac{d\omega}{\pi}\frac{\Gamma(\omega)}{\epsilon-\omega\pm i0^{+}}
\end{split}
\eeS
and 
\be
\label{SE_L}
\begin{split}
\Sigma^{<}_{f\sigma}(\epsilon)&=\langle\sigma^{x}\rangle^{2}_{st}\frac{2}{V}\sum_{k}v^{2}_{k}G^{<}_{11\sigma}(\epsilon,k)\\
&=4\langle\sigma^{x}\rangle^{2}_{st}i\Gamma(\epsilon)f_{\text{neq}}(\epsilon).
\end{split}
\ee
\paragraph{Expectation values}
 The average occupation on the quantum dot (\ref{effective_couplings_1n}) follows from Eqs. (\ref{<_Dyson}) and (\ref{SE_L}).
The average value of the hybridization (\ref{effective_coupling1}) involves the lesser component of the mixed Green's function:
\be
\label{Mixed_G}
G^{<}_{1kf\sigma}=\sqrt{\frac{2}{V}}v_{k}\langle\sigma^{x}\rangle_{st}\left[\hat{G}_{11k\sigma}\cdot \hat{G}_{f\sigma}\right]^{<}.
\ee
Thus,
\be
\label{T_1}
\langle T\rangle_{st}=\frac{2}{\langle\sigma^{x}\rangle_{st}}\sum_{\sigma}\int\frac{d\epsilon}{2\pi}\text{Im}\left[\hat{\Sigma}_{f\sigma}(\epsilon)\cdot\hat{G}_{f\sigma}(\epsilon)\right]^{<}.
\ee
By using Eqs. (\ref{RA_Dyson}), (\ref{<_Dyson}) and (\ref{SE_L}) we readily obtain Eq. (\ref{effective_couplings_1T}) reported in the main text. 
Finally, we briefly derive the expression for the low-energy contribution to the current average value Eq. (\ref{I_f}). In this case the mixed Green's function involved is $G^{<}_{2kf\sigma}(t,t)=i\langle f^{\dagger}_{\sigma}(t)c_{2k\sigma}(t)\rangle_{st}$ and its Dyson's equation reads:
\beS
\label{2kf_current}
G^{<}_{2kf\sigma}(\epsilon)=\sqrt{\frac{2}{V}}v_{k}\langle\sigma^{x}\rangle_{st}G_{21k\sigma}^{<}(\epsilon)G_{f\sigma}^{A}(\epsilon).
\eeS
The average value of the current is:
\be
\label{I_mf1}
\langle I_{f}\rangle_{st}=\sum_{\sigma}\int \frac{d\epsilon}{2\pi}\text{Re}\left[\Sigma^{<}_{21\sigma}(\epsilon)G^{A}_{f\sigma}(\epsilon)\right] ,
\ee 
where 
\beS
\label{Sigma_21}
\begin{split}
\Sigma^{<}_{21\sigma}(\epsilon)&=\langle\sigma^{x}\rangle^{2}_{st}\frac{2}{V}\sum_{k}v^{2}_{k}G_{21k\sigma}^{<}(\epsilon)\\
&=4\langle\sigma^{x}\rangle^{2}_{st}i\Gamma(\epsilon)\frac{f_{L}(\epsilon)-f_{R}(\epsilon)}{2}.
\end{split}
\eeS
In the WBL Eq. (\ref{I_mf1}) gives Eq. (\ref{I_f}). 
\paragraph{$\psi$ fermion Green's function} The Dyson's equation for the $\psi$ fermion reads:
\be
\label{psi_G}
\hat{G}_{\psi}=\hat{G}^{0}_{\psi}+\hat{G}^{0}_{\psi}\cdot\hat{\Sigma}_{\psi}\cdot\hat{G}_{\psi},
\ee  
where the Hartee-Fock self-energy, depicted in Fig. \ref{Lines} c) is: 
\beS
\label{SE_psi}
\hat{\Sigma}_{\psi}=\sigma^{x}\langle T\rangle_{st}, 
\eeS  
In Eq. (\ref{psi_G}) we are using the same notation introduced in Eq. (\ref{Dyson_fRA}), where the hat refers to the matrix structure (\ref{Matrix_Keldysh}).
By performing straightforward calculations we obtain: 
\beS
\label{psiRA_G}
G^{R(A)}_{\psi}(\epsilon)=\sum_{\mu}\sigma^{\mu}G^{R(A)}_{\psi\mu}(\epsilon),
\eeS 
where $\mu=0$ denotes the identity and $\mu=1,2,3$ the remaining Pauli matrices, while $G^{R(A)}_{\psi2}(\epsilon)=0$ and
\beS
\label{psiRA_Gc}
\begin{split}
G^{R(A)}_{\psi0}(\epsilon)&=\frac{1}{2}\left(\frac{1}{\epsilon+\omega_{0}/2\pm i0^{+}}+\frac{1}{\epsilon-\omega_{0}/2\pm i0^{+}}\right),\\
G^{R(A)}_{\psi1}(\epsilon)&=\frac{\sin\theta}{2}\left(\frac{1}{\epsilon+\omega_{0}/2\pm i0^{+}}-\frac{1}{\epsilon-\omega_{0}/2\pm i0^{+}}\right),\\
G^{R(A)}_{\psi3}(\epsilon)&=\frac{\cos\theta}{2}\left(\frac{1}{\epsilon+\omega_{0}/2\pm i0^{+}}-\frac{1}{\epsilon-\omega_{0}/2\pm i0^{+}}\right),
\end{split}
\eeS 
with $\omega_{0}=U\sqrt{1+16\langle T\rangle_{st}^{2}/U^{2}}/2$ and $\theta$ solution of Eq. (\ref{mf_SCE}). Finally, we report the lesser component:
\beS
\label{psiL_G}
G^{<}_{\psi}(\epsilon)=\sum_{\mu}\sigma^{\mu}G^{<}_{\psi\mu}(\epsilon),
\eeS 
where $G^{<}_{\psi2}(\epsilon)=0$ and 
\beS
\label{psiRA_Gc}
\begin{split}
G^{<}_{\psi0}(\epsilon)&=i\pi f(\epsilon)\left[\delta(\epsilon+\omega_{0}/2)+\delta(\epsilon-\omega_{0}/2)\right],\\
G^{<}_{\psi1}(\epsilon)&=i\pi f(\epsilon)\sin\theta\left[\delta(\epsilon+\omega_{0}/2)-\delta(\epsilon-\omega_{0}/2)\right],\\
G^{<}_{\psi3}(\epsilon)&=i\pi f(\epsilon)\cos\theta\left[\delta(\epsilon+\omega_{0}/2)-\delta(\epsilon-\omega_{0}/2)\right].
\end{split}
\eeS 

\section{RPA corrections to the spin correlation function}
\label{RPA_eqs}

In this section, we compute the RPA correction to the $\sigma^{x}$ mode, which describes valence fluctuations on the impurity site. In terms of the fermionic representation introduced in Eq. (\ref{Abrikosov_Sf}) the bare $\Pi_{xx}$ propagator reads:
\beS
\label{bare_xx}
\hat{\Pi}^{0}_{xx}(t,t^\prime)=-i\text{Tr}\left[\sigma^{x}\hat{G}_{\psi}(t,t^{\prime})\sigma^{x}\hat{G}_{\psi}(t^{\prime},t)\right],
\eeS
where $\hat{G}_{\psi}$ is the Hartree-Fock $\psi$ fermion Green's function in Eq. (\ref{psi_G}).
As shown in Fig. \ref{diagrammi_transport} c) the Dyson's equation reads:
\beS
\label{Dyson_bose}
\hat{\Pi}_{xx}=\hat{\Pi}_{xx}^{0}+\hat{\Pi}_{xx}^{0}\cdot\hat{\Sigma}_{xx}\cdot\hat{\Pi}_{xx},
\eeS
where we adopt the notation introduced in Eq. (\ref{Matrix_Keldysh}). At RPA level the bosonic self-energy reads: 
\be
\label{self_bosons}
\hat{\Sigma}_{xx}=\hat{\chi}_{TT},
\ee
with: 
\beS
\label{TT}
\chi_{TT}(t,t^{\prime})=-i\langle T_{\mathcal{C}}(\delta T(t)\delta T(t^{\prime}))\rangle
\eeS
where $\delta T=T-\langle T\rangle_{st}$, and $T$ is the hybridization operator in Eq. (\ref{Op_Hybridisation}).
Within the WBL, introduced in Eq. (\ref{f_self_e_WBL}), the evaluation of the bosonic self-energy (\ref{self_bosons}) is considerably simplified. We find: 
\beS
\label{TT_L}
\begin{split}
\chi^{<}_{TT}(\omega)&=-i\frac{1}{\pi\langle\sigma^{x}\rangle^{2}_{st}}\int d\epsilon\Big[G^{<}_{f}(\epsilon+\omega)\Sigma_{f}^{>}(\epsilon)\\
&+\Sigma^{<}_{f}(\epsilon+\omega)G^{>}_{f}(\epsilon)\Big]\\
&-i\frac{2}{\pi\langle\sigma^{x}\rangle^{2}_{st}}\int d\epsilon\Sigma^{<}_{f}(\epsilon+\omega)\Sigma^{>}_{f}(\epsilon)\\
&\text{Re}\Big[G^{R}_{f}(\epsilon+\omega)G^{R}_{f}(\epsilon)\Big],
\end{split}
\eeS
and 
\beS
\label{TT_R}
\begin{split}
\chi^{R}_{TT}(\omega)&=-i\frac{1}{\pi\langle\sigma^{x}\rangle^{2}_{st}}\int d\epsilon\Sigma^{<}_{f}(\epsilon)\Big[G^{R}_{f}(\epsilon+\omega)\\
&+G^{A}_{f}(\epsilon-\omega)\Big]\\
&-i\frac{2\Sigma^{R}}{\pi\langle\sigma^{x}\rangle^{2}_{st}}\int d\epsilon\Sigma^{<}_{f}(\epsilon)\\
&\Big[G^{R}_{f}(\epsilon+\omega)G^{R}_{f}(\epsilon)-G^{A}_{f}(\epsilon-\omega)G^{A}_{f}(\epsilon)\Big].
\end{split}
\eeS
  
 \section{Transient dynamics of the effective Resonant level model}
\label{K_RLM_transient}

The dynamics of the spin degree of freedom is influenced by the time-dependent expectation value of the hybridization Eq. (\ref{T_2}). By assuming a slowly varying electrochemical potential (\ref{electrochemical_pot}), we compute Eq. (\ref{T_2}) to the first-order in the gradient expansion Eq. (\ref{gradient_exp}).
To this aim we define the Wigner transform of the $f_{\sigma}$ pseudofermion Green's function:
\beS
\label{Wigner_T}
G^{R(A)}_{f\sigma}(t,\epsilon)=\int d\tau e^{i\epsilon\tau} G^{R(A)}_{f\sigma}\left(t+\frac{\tau}{2},t-\frac{\tau}{2}\right),  
\eeS
which satisfies the Dyson's equation: 
\beS
\label{RA_Dyson_t}
\left(\epsilon-\Sigma^{R(A)}_{f\sigma}(t,\epsilon)\right)\star G^{R(A)}_{f\sigma}(t,\epsilon)=1
\eeS
where $\star$ denotes the Moyal product introduced in the main text. The solution of the Dyson's equation  up to first-order is:
\beS
\label{gEXPANSION1}
G^{R(A)}_{f\sigma}(t,\epsilon)=\frac{1}{\epsilon-\Sigma_{f\sigma}^{R(A)}(t,\epsilon)}
\eeS
where in the WBL the time-dependent self-energy is $\Sigma_{f\sigma}^{R(A)}(t,\epsilon)=\mp2i\Gamma_{0}\langle\sigma^{x}(t)\rangle^2$. 
Instead, the lesser self-energy is given by:
\be
\label{SE_L_t}
\begin{split}
\Sigma^{<}_{f\sigma}(t,\epsilon)&=2i\Gamma_{0}\langle\sigma^{x}(t)\rangle^2-\frac{2\Gamma_{0}}{\pi}\int d\tau\frac{e^{i\epsilon\tau}}{\tau}\\
&\cos\gamma(t,\tau)\Big\langle\sigma^{x}\left(t+\frac{\tau}{2}\right)\Big\rangle\Big\langle\sigma^{x}\left(t-\frac{\tau}{2}\right)\Big\rangle\\
&\simeq4\Gamma_{0}i\langle\sigma^{x}(t)\rangle^2f_{neq}(t,\epsilon),
\end{split}
\ee
where $\gamma(t,\tau)=\int^{t+\tau/2}_{t-\tau/2}\mu_{L}(x)dx$ and the nonequilibrium distribution reads 
\beS
\label{fneq_t}
f_{neq}(t,\epsilon)=\frac{1}{2}+\frac{i}{2\pi}\int d\tau\frac{e^{i\epsilon\tau}}{\tau}\cos\gamma(t,\tau).
\eeS
In the last passage of Eq. (\ref{SE_L_t}), we assume that the dependence of $\langle\sigma^{x}(t)\rangle$ on the relative time $\tau$ is negligible. 

In the following, we report the zeroth and first-order contributions to the gradient expansion of $\langle T(t)\rangle$.
\paragraph{Zeroth order}
The zeroth order contribution, first term in Eq. (\ref{gradient_exp}), reads: 
\beS
\label{T_zeroO}
\langle T(t)\rangle^{(0)}=\frac{4}{\langle\sigma^{x}(t)\rangle}\int d\epsilon A_{f}(t,\epsilon)\epsilon f_{\text{neq}}(t,\epsilon),
\eeS
where the $f_{\sigma}$ pseudofermion time-dependent spectral function is  
\beS
\label{td_spectral}
A_{f}(t,\epsilon)=\frac{1}{\pi}\frac{\Gamma(t)}{\epsilon^2+\Gamma(t)^{2}},
\eeS
with $\Gamma(t)=2\Gamma_{0}\langle\sigma^{x}(t)\rangle^{2}$.
\paragraph{First order} 
The first order correction to the quasistatic approximation is the second term of Eq. (\ref{gradient_exp}), which reads:
\beS
\label{T_first1}
\begin{split}
\langle T(t)\rangle^{(1)}&=\frac{1}{\pi\langle\sigma^{x}(t)\rangle}\text{Im}\int d\epsilon \Big[i\Big(\partial_{\epsilon}\Sigma^{<}_{f}(t,\epsilon)\partial_{t}\Sigma^{A}_{f}(t,\epsilon)\\
&+\partial_{t}\Sigma^{<}_{f}(t,\epsilon)\Big)G^{A}_{f}(t,\epsilon)^{2}\Big].
\end{split}
\eeS 
After straightforward calculations we obtain
\beS
\label{T_first2}
\begin{split}
\langle T(t)\rangle^{(1)}&=-\frac{2\Gamma(t)}{\pi\langle\sigma^{x}(t)\rangle}\int d\epsilon \Big[\text{Im}\left[G^{A}_{f}(t,\epsilon)^{2}\right]\partial_{t}f_{\text{neq}}(t,\epsilon)\\
&+2\frac{\partial_{t}\langle\sigma^{x}(t)\rangle}{\langle\sigma^{x}(t)\rangle}\Big(f_{\text{neq}}(t,\epsilon)\text{Im}\left[G^{A}_{f}(t,\epsilon)^{2}\right]\\
&+\partial_{\epsilon}f_{\text{neq}}(t,\epsilon)\Gamma(t)\text{Re}\left[G^{A}_{f}(t,\epsilon)^{2}\right]\Big)\Big].
\end{split}
\eeS 
Since $\partial_{t}\langle\sigma^{x}(t)\rangle=U\langle\sigma^{y}(t)\rangle/2$ the latter contribution modifies the Heisenberg equation (\ref{Ising_dyn_mu}) by introducing a finite relaxation in the evolution of the $\langle\sigma^{y}(t)\rangle$ component.

\vspace{0.2cm}

\bibliographystyle{apsrev}
\bibliography{mybiblio}

\begin{thebibliography}{49}
\expandafter\ifx\csname natexlab\endcsname\relax\def\natexlab#1{#1}\fi
\expandafter\ifx\csname bibnamefont\endcsname\relax
  \def\bibnamefont#1{#1}\fi
\expandafter\ifx\csname bibfnamefont\endcsname\relax
  \def\bibfnamefont#1{#1}\fi
\expandafter\ifx\csname citenamefont\endcsname\relax
  \def\citenamefont#1{#1}\fi
\expandafter\ifx\csname url\endcsname\relax
  \def\url#1{\texttt{#1}}\fi
\expandafter\ifx\csname urlprefix\endcsname\relax\def\urlprefix{URL }\fi
\providecommand{\bibinfo}[2]{#2}
\providecommand{\eprint}[2][]{\url{#2}}

\bibitem[{\citenamefont{Hewson}(1993)}]{hewson_1993}
\bibinfo{author}{\bibfnamefont{A.~C.} \bibnamefont{Hewson}},
  \emph{\bibinfo{title}{The Kondo Problem to Heavy Fermions}}, Cambridge
  Studies in Magnetism (\bibinfo{publisher}{Cambridge University Press},
  \bibinfo{year}{1993}).

\bibitem[{\citenamefont{Kondo}(1964)}]{Kondo_64}
\bibinfo{author}{\bibfnamefont{J.}~\bibnamefont{Kondo}},
  \bibinfo{journal}{Progress of Theoretical Physics}
  \textbf{\bibinfo{volume}{32}}, \bibinfo{pages}{37} (\bibinfo{year}{1964}),
  \urlprefix\url{http://dx.doi.org/10.1143/PTP.32.37}.

\bibitem[{\citenamefont{Anderson}(1970)}]{Poor_man}
\bibinfo{author}{\bibfnamefont{P.~W.} \bibnamefont{Anderson}},
  \bibinfo{journal}{Journal of Physics C: Solid State Physics}
  \textbf{\bibinfo{volume}{3}}, \bibinfo{pages}{2436} (\bibinfo{year}{1970}),
  \urlprefix\url{http://stacks.iop.org/0022-3719/3/i=12/a=008}.

\bibitem[{\citenamefont{Stefan et~al.}(2014)\citenamefont{Stefan,
  Mart\'{\i}nez-Blanco, Yang, Kanisawa, and Erwin}}]{Stefan_2014}
\bibinfo{author}{\bibfnamefont{F.}~\bibnamefont{Stefan}},
  \bibinfo{author}{\bibfnamefont{J.}~\bibnamefont{Mart\'{\i}nez-Blanco}},
  \bibinfo{author}{\bibfnamefont{J.}~\bibnamefont{Yang}},
  \bibinfo{author}{\bibfnamefont{K.}~\bibnamefont{Kanisawa}}, \bibnamefont{and}
  \bibinfo{author}{\bibfnamefont{S.~C.} \bibnamefont{Erwin}},
  \bibinfo{journal}{Nature Nanotechnology} \textbf{\bibinfo{volume}{9}},
  \bibinfo{pages}{505} (\bibinfo{year}{2014}),
  \urlprefix\url{https://doi.org/10.1038/nnano.2014.129}.

\bibitem[{\citenamefont{Mart\'{\i}nez-Blanco
  et~al.}(2015)\citenamefont{Mart\'{\i}nez-Blanco, Nacci, Erwin, Kanisawa,
  Locane, Thomas, von Oppen, Brouwer, and Stefan}}]{Stefan_2015}
\bibinfo{author}{\bibfnamefont{J.}~\bibnamefont{Mart\'{\i}nez-Blanco}},
  \bibinfo{author}{\bibfnamefont{C.}~\bibnamefont{Nacci}},
  \bibinfo{author}{\bibfnamefont{S.~C.} \bibnamefont{Erwin}},
  \bibinfo{author}{\bibfnamefont{K.}~\bibnamefont{Kanisawa}},
  \bibinfo{author}{\bibfnamefont{E.}~\bibnamefont{Locane}},
  \bibinfo{author}{\bibfnamefont{M.}~\bibnamefont{Thomas}},
  \bibinfo{author}{\bibfnamefont{F.}~\bibnamefont{von Oppen}},
  \bibinfo{author}{\bibfnamefont{P.~W.} \bibnamefont{Brouwer}},
  \bibnamefont{and} \bibinfo{author}{\bibfnamefont{F.}~\bibnamefont{Stefan}},
  \bibinfo{journal}{Nature Physics} \textbf{\bibinfo{volume}{11}},
  \bibinfo{pages}{640} (\bibinfo{year}{2015}),
  \urlprefix\url{https://doi.org/10.1038/nphys3385}.

\bibitem[{\citenamefont{Pan et~al.}(2015)\citenamefont{Pan, Yang, Erwin,
  Kanisawa, and F\"olsch}}]{Stefan_2015_PRL}
\bibinfo{author}{\bibfnamefont{Y.}~\bibnamefont{Pan}},
  \bibinfo{author}{\bibfnamefont{J.}~\bibnamefont{Yang}},
  \bibinfo{author}{\bibfnamefont{S.~C.} \bibnamefont{Erwin}},
  \bibinfo{author}{\bibfnamefont{K.}~\bibnamefont{Kanisawa}}, \bibnamefont{and}
  \bibinfo{author}{\bibfnamefont{S.}~\bibnamefont{F\"olsch}},
  \bibinfo{journal}{Phys. Rev. Lett.} \textbf{\bibinfo{volume}{115}},
  \bibinfo{pages}{076803} (\bibinfo{year}{2015}),
  \urlprefix\url{https://link.aps.org/doi/10.1103/PhysRevLett.115.076803}.

\bibitem[{\citenamefont{Kastner}(1993)}]{Kastner_PT}
\bibinfo{author}{\bibfnamefont{M.~A.} \bibnamefont{Kastner}},
  \bibinfo{journal}{Physics Today} \textbf{\bibinfo{volume}{46}},
  \bibinfo{pages}{24} (\bibinfo{year}{1993}),
  \urlprefix\url{https://doi.org/10.1063/1.881393}.

\bibitem[{\citenamefont{Ashoori}(1996)}]{Ashoori_Nat}
\bibinfo{author}{\bibfnamefont{R.~C.} \bibnamefont{Ashoori}},
  \bibinfo{journal}{Nature} \textbf{\bibinfo{volume}{379}},
  \bibinfo{pages}{413} (\bibinfo{year}{1996}),
  \urlprefix\url{https://doi.org/10.1038/379413a0}.

\bibitem[{\citenamefont{Goldhaber-Gordon
  et~al.}(1998)\citenamefont{Goldhaber-Gordon, G\"ores, Kastner, Shtrikman,
  Mahalu, and Meirav}}]{GoldhaberGordon_prl98}
\bibinfo{author}{\bibfnamefont{D.}~\bibnamefont{Goldhaber-Gordon}},
  \bibinfo{author}{\bibfnamefont{J.}~\bibnamefont{G\"ores}},
  \bibinfo{author}{\bibfnamefont{M.~A.} \bibnamefont{Kastner}},
  \bibinfo{author}{\bibfnamefont{H.}~\bibnamefont{Shtrikman}},
  \bibinfo{author}{\bibfnamefont{D.}~\bibnamefont{Mahalu}}, \bibnamefont{and}
  \bibinfo{author}{\bibfnamefont{U.}~\bibnamefont{Meirav}},
  \bibinfo{journal}{Phys. Rev. Lett.} \textbf{\bibinfo{volume}{81}},
  \bibinfo{pages}{5225} (\bibinfo{year}{1998}).

\bibitem[{\citenamefont{Cronenwett et~al.}(1998)\citenamefont{Cronenwett,
  Oosterkamp, and Kouwenhoven}}]{Kouwenhoven_science98}
\bibinfo{author}{\bibfnamefont{S.~M.} \bibnamefont{Cronenwett}},
  \bibinfo{author}{\bibfnamefont{T.~H.} \bibnamefont{Oosterkamp}},
  \bibnamefont{and} \bibinfo{author}{\bibfnamefont{L.~P.}
  \bibnamefont{Kouwenhoven}}, \bibinfo{journal}{Science}
  \textbf{\bibinfo{volume}{281}}, \bibinfo{pages}{540} (\bibinfo{year}{1998}),
  \eprint{http://www.sciencemag.org/cgi/reprint/281/5376/540.pdf},
  \urlprefix\url{http://www.sciencemag.org/cgi/content/abstract/281/5376/540}.

\bibitem[{\citenamefont{Glazman and Raikh}(1988)}]{Glazman88}
\bibinfo{author}{\bibfnamefont{L.~I.} \bibnamefont{Glazman}} \bibnamefont{and}
  \bibinfo{author}{\bibfnamefont{M.~E.} \bibnamefont{Raikh}},
  \bibinfo{journal}{JETP Lett.} \textbf{\bibinfo{volume}{47}},
  \bibinfo{pages}{452} (\bibinfo{year}{1988}).

\bibitem[{\citenamefont{Ng and Lee}(1988)}]{NgLee88}
\bibinfo{author}{\bibfnamefont{T.~K.} \bibnamefont{Ng}} \bibnamefont{and}
  \bibinfo{author}{\bibfnamefont{P.~A.} \bibnamefont{Lee}},
  \bibinfo{journal}{Phys. Rev. Lett.} \textbf{\bibinfo{volume}{61}},
  \bibinfo{pages}{1768} (\bibinfo{year}{1988}).

\bibitem[{\citenamefont{Kobayashi et~al.}(2010)\citenamefont{Kobayashi,
  Tsuruta, Sasaki, Fujisawa, Tokura, and Akazaki}}]{Kobayashi_10}
\bibinfo{author}{\bibfnamefont{T.}~\bibnamefont{Kobayashi}},
  \bibinfo{author}{\bibfnamefont{S.}~\bibnamefont{Tsuruta}},
  \bibinfo{author}{\bibfnamefont{S.}~\bibnamefont{Sasaki}},
  \bibinfo{author}{\bibfnamefont{T.}~\bibnamefont{Fujisawa}},
  \bibinfo{author}{\bibfnamefont{Y.}~\bibnamefont{Tokura}}, \bibnamefont{and}
  \bibinfo{author}{\bibfnamefont{T.}~\bibnamefont{Akazaki}},
  \bibinfo{journal}{Phys. Rev. Lett.} \textbf{\bibinfo{volume}{104}},
  \bibinfo{pages}{036804} (\bibinfo{year}{2010}),
  \urlprefix\url{https://link.aps.org/doi/10.1103/PhysRevLett.104.036804}.

\bibitem[{\citenamefont{Anders and Schiller}(2005)}]{tnrg_Anders}
\bibinfo{author}{\bibfnamefont{F.~B.} \bibnamefont{Anders}} \bibnamefont{and}
  \bibinfo{author}{\bibfnamefont{A.}~\bibnamefont{Schiller}},
  \bibinfo{journal}{Phys. Rev. Lett.} \textbf{\bibinfo{volume}{95}},
  \bibinfo{pages}{196801} (\bibinfo{year}{2005}),
  \urlprefix\url{http://link.aps.org/abstract/PRL/v95/e196801}.

\bibitem[{\citenamefont{Anders and Schiller}(2006)}]{Anders_prb06}
\bibinfo{author}{\bibfnamefont{F.~B.} \bibnamefont{Anders}} \bibnamefont{and}
  \bibinfo{author}{\bibfnamefont{A.}~\bibnamefont{Schiller}},
  \bibinfo{journal}{Phys. Rev. B} \textbf{\bibinfo{volume}{74}},
  \bibinfo{pages}{245113} (\bibinfo{year}{2006}).

\bibitem[{\citenamefont{Anders}(2008)}]{Andersprl2008}
\bibinfo{author}{\bibfnamefont{F.~B.} \bibnamefont{Anders}},
  \bibinfo{journal}{Phys. Rev. Lett.} \textbf{\bibinfo{volume}{101}},
  \bibinfo{pages}{066804} (\bibinfo{year}{2008}),
  \urlprefix\url{https://link.aps.org/doi/10.1103/PhysRevLett.101.066804}.

\bibitem[{\citenamefont{Werner et~al.}(2009)\citenamefont{Werner, Oka, and
  Millis}}]{Werner_Keldysh_09}
\bibinfo{author}{\bibfnamefont{P.}~\bibnamefont{Werner}},
  \bibinfo{author}{\bibfnamefont{T.}~\bibnamefont{Oka}}, \bibnamefont{and}
  \bibinfo{author}{\bibfnamefont{A.~J.} \bibnamefont{Millis}},
  \bibinfo{journal}{Phys. Rev. B} \textbf{\bibinfo{volume}{79}},
  \bibinfo{pages}{035320} (\bibinfo{year}{2009}).

\bibitem[{\citenamefont{Schir\'o and Fabrizio}(2009)}]{Marco_2009}
\bibinfo{author}{\bibfnamefont{M.}~\bibnamefont{Schir\'o}} \bibnamefont{and}
  \bibinfo{author}{\bibfnamefont{M.}~\bibnamefont{Fabrizio}},
  \bibinfo{journal}{Phys. Rev. B} \textbf{\bibinfo{volume}{79}},
  \bibinfo{pages}{153302} (\bibinfo{year}{2009}),
  \urlprefix\url{https://link.aps.org/doi/10.1103/PhysRevB.79.153302}.

\bibitem[{\citenamefont{White and Feiguin}(2004)}]{Whiteprl2004}
\bibinfo{author}{\bibfnamefont{S.~R.} \bibnamefont{White}} \bibnamefont{and}
  \bibinfo{author}{\bibfnamefont{A.~E.} \bibnamefont{Feiguin}},
  \bibinfo{journal}{Phys. Rev. Lett.} \textbf{\bibinfo{volume}{93}},
  \bibinfo{pages}{076401} (\bibinfo{year}{2004}),
  \urlprefix\url{https://link.aps.org/doi/10.1103/PhysRevLett.93.076401}.

\bibitem[{\citenamefont{Schmitteckert}(2004)}]{Schmitteckertprb2004}
\bibinfo{author}{\bibfnamefont{P.}~\bibnamefont{Schmitteckert}},
  \bibinfo{journal}{Phys. Rev. B} \textbf{\bibinfo{volume}{70}},
  \bibinfo{pages}{121302(R)} (\bibinfo{year}{2004}),
  \urlprefix\url{https://link.aps.org/doi/10.1103/PhysRevB.70.121302}.

\bibitem[{\citenamefont{Boulat et~al.}(2008)\citenamefont{Boulat, Saleur, and
  Schmitteckert}}]{Saleur08}
\bibinfo{author}{\bibfnamefont{E.}~\bibnamefont{Boulat}},
  \bibinfo{author}{\bibfnamefont{H.}~\bibnamefont{Saleur}}, \bibnamefont{and}
  \bibinfo{author}{\bibfnamefont{P.}~\bibnamefont{Schmitteckert}},
  \bibinfo{journal}{Phys. Rev. Lett.} \textbf{\bibinfo{volume}{101}},
  \bibinfo{pages}{140601} (\bibinfo{year}{2008}).

\bibitem[{\citenamefont{Kehrein}(2005)}]{Kehrein_Kondo}
\bibinfo{author}{\bibfnamefont{S.}~\bibnamefont{Kehrein}},
  \bibinfo{journal}{Phys. Rev. Lett.} \textbf{\bibinfo{volume}{95}},
  \bibinfo{pages}{056602} (\bibinfo{year}{2005}),
  \urlprefix\url{https://link.aps.org/doi/10.1103/PhysRevLett.95.056602}.

\bibitem[{\citenamefont{Fritsch and Kehrein}(2010)}]{Kehrein_magneticfield}
\bibinfo{author}{\bibfnamefont{P.}~\bibnamefont{Fritsch}} \bibnamefont{and}
  \bibinfo{author}{\bibfnamefont{S.}~\bibnamefont{Kehrein}},
  \bibinfo{journal}{Phys. Rev. B} \textbf{\bibinfo{volume}{81}},
  \bibinfo{pages}{035113} (\bibinfo{year}{2010}).

\bibitem[{\citenamefont{Tomaras and Kehrein}(2011)}]{Tomaras_11}
\bibinfo{author}{\bibfnamefont{C.}~\bibnamefont{Tomaras}} \bibnamefont{and}
  \bibinfo{author}{\bibfnamefont{S.}~\bibnamefont{Kehrein}},
  \bibinfo{journal}{EPL (Europhysics Letters)} \textbf{\bibinfo{volume}{93}},
  \bibinfo{pages}{47011} (\bibinfo{year}{2011}),
  \urlprefix\url{http://stacks.iop.org/0295-5075/93/i=4/a=47011}.

\bibitem[{\citenamefont{Nordlander et~al.}(1999)\citenamefont{Nordlander,
  Pustilnik, Meir, Wingreen, and Langreth}}]{Pustilnik_1999}
\bibinfo{author}{\bibfnamefont{P.}~\bibnamefont{Nordlander}},
  \bibinfo{author}{\bibfnamefont{M.}~\bibnamefont{Pustilnik}},
  \bibinfo{author}{\bibfnamefont{Y.}~\bibnamefont{Meir}},
  \bibinfo{author}{\bibfnamefont{N.~S.} \bibnamefont{Wingreen}},
  \bibnamefont{and} \bibinfo{author}{\bibfnamefont{D.~C.}
  \bibnamefont{Langreth}}, \bibinfo{journal}{Phys. Rev. Lett.}
  \textbf{\bibinfo{volume}{83}}, \bibinfo{pages}{808} (\bibinfo{year}{1999}),
  \urlprefix\url{https://link.aps.org/doi/10.1103/PhysRevLett.83.808}.

\bibitem[{\citenamefont{Kaminski et~al.}(2000)\citenamefont{Kaminski, Nazarov,
  and Glazman}}]{Glazman00}
\bibinfo{author}{\bibfnamefont{A.}~\bibnamefont{Kaminski}},
  \bibinfo{author}{\bibfnamefont{Y.~V.} \bibnamefont{Nazarov}},
  \bibnamefont{and} \bibinfo{author}{\bibfnamefont{L.~I.}
  \bibnamefont{Glazman}}, \bibinfo{journal}{Phys. Rev. B}
  \textbf{\bibinfo{volume}{62}}, \bibinfo{pages}{8154} (\bibinfo{year}{2000}).

\bibitem[{\citenamefont{Rosch et~al.}(2003)\citenamefont{Rosch, Paaske, Kroha,
  and W\"olfle}}]{Rosch_magneticfield}
\bibinfo{author}{\bibfnamefont{A.}~\bibnamefont{Rosch}},
  \bibinfo{author}{\bibfnamefont{J.}~\bibnamefont{Paaske}},
  \bibinfo{author}{\bibfnamefont{J.}~\bibnamefont{Kroha}}, \bibnamefont{and}
  \bibinfo{author}{\bibfnamefont{P.}~\bibnamefont{W\"olfle}},
  \bibinfo{journal}{Phys. Rev. Lett.} \textbf{\bibinfo{volume}{90}},
  \bibinfo{pages}{076804} (\bibinfo{year}{2003}).

\bibitem[{\citenamefont{Metzner et~al.}(2012)\citenamefont{Metzner, Salmhofer,
  Honerkamp, Meden, and Sch\"onhammer}}]{Metzner_2012}
\bibinfo{author}{\bibfnamefont{W.}~\bibnamefont{Metzner}},
  \bibinfo{author}{\bibfnamefont{M.}~\bibnamefont{Salmhofer}},
  \bibinfo{author}{\bibfnamefont{C.}~\bibnamefont{Honerkamp}},
  \bibinfo{author}{\bibfnamefont{V.}~\bibnamefont{Meden}}, \bibnamefont{and}
  \bibinfo{author}{\bibfnamefont{K.}~\bibnamefont{Sch\"onhammer}},
  \bibinfo{journal}{Rev. Mod. Phys.} \textbf{\bibinfo{volume}{84}},
  \bibinfo{pages}{299} (\bibinfo{year}{2012}),
  \urlprefix\url{https://link.aps.org/doi/10.1103/RevModPhys.84.299}.

\bibitem[{\citenamefont{Schoeller}(2009)}]{Schoeller2009}
\bibinfo{author}{\bibfnamefont{H.}~\bibnamefont{Schoeller}},
  \bibinfo{journal}{The European Physical Journal Special Topics}
  \textbf{\bibinfo{volume}{168}}, \bibinfo{pages}{179} (\bibinfo{year}{2009}),
  ISSN \bibinfo{issn}{1951-6401},
  \urlprefix\url{https://doi.org/10.1140/epjst/e2009-00962-3}.

\bibitem[{\citenamefont{Ashida et~al.}(2018)\citenamefont{Ashida, Shi,
  Ba\~nuls, Cirac, and Demler}}]{Cirac_QI}
\bibinfo{author}{\bibfnamefont{Y.}~\bibnamefont{Ashida}},
  \bibinfo{author}{\bibfnamefont{T.}~\bibnamefont{Shi}},
  \bibinfo{author}{\bibfnamefont{M.~C.} \bibnamefont{Ba\~nuls}},
  \bibinfo{author}{\bibfnamefont{J.~I.} \bibnamefont{Cirac}}, \bibnamefont{and}
  \bibinfo{author}{\bibfnamefont{E.}~\bibnamefont{Demler}},
  \bibinfo{journal}{Phys. Rev. Lett.} \textbf{\bibinfo{volume}{121}},
  \bibinfo{pages}{026805} (\bibinfo{year}{2018}),
  \urlprefix\url{https://link.aps.org/doi/10.1103/PhysRevLett.121.026805}.

\bibitem[{\citenamefont{Lanat\`a and Strand}(2012)}]{Nicola_Hugo}
\bibinfo{author}{\bibfnamefont{N.}~\bibnamefont{Lanat\`a}} \bibnamefont{and}
  \bibinfo{author}{\bibfnamefont{H.~U.~R.} \bibnamefont{Strand}},
  \bibinfo{journal}{Phys. Rev. B} \textbf{\bibinfo{volume}{86}},
  \bibinfo{pages}{115310} (\bibinfo{year}{2012}),
  \urlprefix\url{https://link.aps.org/doi/10.1103/PhysRevB.86.115310}.

\bibitem[{\citenamefont{Citro and Romeo}(2016)}]{CitroSB}
\bibinfo{author}{\bibfnamefont{R.}~\bibnamefont{Citro}} \bibnamefont{and}
  \bibinfo{author}{\bibfnamefont{F.}~\bibnamefont{Romeo}},
  \bibinfo{journal}{Journal of Physics: Conference Series}
  \textbf{\bibinfo{volume}{696}}, \bibinfo{pages}{012014}
  (\bibinfo{year}{2016}),
  \urlprefix\url{http://stacks.iop.org/1742-6596/696/i=1/a=012014}.

\bibitem[{\citenamefont{Ludovico and Capone}(2018)}]{Florencia_Qd}
\bibinfo{author}{\bibfnamefont{M.~F.} \bibnamefont{Ludovico}} \bibnamefont{and}
  \bibinfo{author}{\bibfnamefont{M.}~\bibnamefont{Capone}},
  \bibinfo{journal}{Phys. Rev. B} \textbf{\bibinfo{volume}{98}},
  \bibinfo{pages}{235409} (\bibinfo{year}{2018}),
  \urlprefix\url{https://link.aps.org/doi/10.1103/PhysRevB.98.235409}.

\bibitem[{\citenamefont{Dong and Lei}(2001)}]{Dong_qd}
\bibinfo{author}{\bibfnamefont{B.}~\bibnamefont{Dong}} \bibnamefont{and}
  \bibinfo{author}{\bibfnamefont{X.~L.} \bibnamefont{Lei}},
  \bibinfo{journal}{Phys. Rev. B} \textbf{\bibinfo{volume}{63}},
  \bibinfo{pages}{235306} (\bibinfo{year}{2001}),
  \urlprefix\url{https://link.aps.org/doi/10.1103/PhysRevB.63.235306}.

\bibitem[{\citenamefont{Raimondi and Schwab}(1999)}]{RAIMONDI_SB}
\bibinfo{author}{\bibfnamefont{R.}~\bibnamefont{Raimondi}} \bibnamefont{and}
  \bibinfo{author}{\bibfnamefont{P.}~\bibnamefont{Schwab}},
  \bibinfo{journal}{Superlattices and Microstructures}
  \textbf{\bibinfo{volume}{25}}, \bibinfo{pages}{1141 } (\bibinfo{year}{1999}),
  ISSN \bibinfo{issn}{0749-6036},
  \urlprefix\url{http://www.sciencedirect.com/science/article/pii/S0749603699907231}.

\bibitem[{\citenamefont{Mehta and Andrei}(2006)}]{Andrei06}
\bibinfo{author}{\bibfnamefont{P.}~\bibnamefont{Mehta}} \bibnamefont{and}
  \bibinfo{author}{\bibfnamefont{N.}~\bibnamefont{Andrei}},
  \bibinfo{journal}{Phys. Rev. Lett.} \textbf{\bibinfo{volume}{96}},
  \bibinfo{pages}{216802} (\bibinfo{year}{2006}).

\bibitem[{\citenamefont{Bolech and Shah}(2016)}]{Bolech2016}
\bibinfo{author}{\bibfnamefont{C.~J.} \bibnamefont{Bolech}} \bibnamefont{and}
  \bibinfo{author}{\bibfnamefont{N.}~\bibnamefont{Shah}},
  \bibinfo{journal}{Phys. Rev. B} \textbf{\bibinfo{volume}{93}},
  \bibinfo{pages}{085441} (\bibinfo{year}{2016}),
  \urlprefix\url{https://link.aps.org/doi/10.1103/PhysRevB.93.085441}.

\bibitem[{\citenamefont{Guerci and Fabrizio}(2017)}]{Daniele-SS}
\bibinfo{author}{\bibfnamefont{D.}~\bibnamefont{Guerci}} \bibnamefont{and}
  \bibinfo{author}{\bibfnamefont{M.}~\bibnamefont{Fabrizio}},
  \bibinfo{journal}{Phys. Rev. B} \textbf{\bibinfo{volume}{96}},
  \bibinfo{pages}{201106(R)} (\bibinfo{year}{2017}),
  \urlprefix\url{https://link.aps.org/doi/10.1103/PhysRevB.96.201106}.

\bibitem[{\citenamefont{Schir\'o and Fabrizio}(2010)}]{Marco-PRL}
\bibinfo{author}{\bibfnamefont{M.}~\bibnamefont{Schir\'o}} \bibnamefont{and}
  \bibinfo{author}{\bibfnamefont{M.}~\bibnamefont{Fabrizio}},
  \bibinfo{journal}{Phys. Rev. Lett.} \textbf{\bibinfo{volume}{105}},
  \bibinfo{pages}{076401} (\bibinfo{year}{2010}),
  \urlprefix\url{http://link.aps.org/doi/10.1103/PhysRevLett.105.076401}.

\bibitem[{\citenamefont{Rammer}(2007)}]{Rammer_book_noneq}
\bibinfo{author}{\bibfnamefont{J.}~\bibnamefont{Rammer}},
  \emph{\bibinfo{title}{Quantum Field Theory of Nonequilibrium States}}
  (\bibinfo{publisher}{Cambridge University Press}, \bibinfo{year}{2007}).

\bibitem[{\citenamefont{Haug and Jauho}(1996)}]{HAUG_JAUHO}
\bibinfo{author}{\bibfnamefont{H.}~\bibnamefont{Haug}} \bibnamefont{and}
  \bibinfo{author}{\bibfnamefont{A.~P.} \bibnamefont{Jauho}},
  \emph{\bibinfo{title}{Quantum Kinetics in Transport and Optics of
  Semiconductors}} (\bibinfo{publisher}{Springer}, \bibinfo{year}{1996}).

\bibitem[{\citenamefont{Arseev}(2015)}]{Arseev_2015}
\bibinfo{author}{\bibfnamefont{P.~I.} \bibnamefont{Arseev}},
  \bibinfo{journal}{Phys. Usp.} \textbf{\bibinfo{volume}{58}},
  \bibinfo{pages}{1159} (\bibinfo{year}{2015}),
  \urlprefix\url{https://ufn.ru/en/articles/2015/12/b/}.

\bibitem[{\citenamefont{Lanat\`a}(2010)}]{Nicola1}
\bibinfo{author}{\bibfnamefont{N.}~\bibnamefont{Lanat\`a}},
  \bibinfo{journal}{Phys. Rev. B} \textbf{\bibinfo{volume}{82}},
  \bibinfo{pages}{195326} (\bibinfo{year}{2010}),
  \urlprefix\url{https://link.aps.org/doi/10.1103/PhysRevB.82.195326}.

\bibitem[{\citenamefont{Baruselli and Fabrizio}(2012)}]{Pierpaolo}
\bibinfo{author}{\bibfnamefont{P.~P.} \bibnamefont{Baruselli}}
  \bibnamefont{and} \bibinfo{author}{\bibfnamefont{M.}~\bibnamefont{Fabrizio}},
  \bibinfo{journal}{Phys. Rev. B} \textbf{\bibinfo{volume}{85}},
  \bibinfo{pages}{073106} (\bibinfo{year}{2012}),
  \urlprefix\url{http://link.aps.org/doi/10.1103/PhysRevB.85.073106}.

\bibitem[{\citenamefont{Abrikosov}(1965)}]{Abrikosov_Fermions}
\bibinfo{author}{\bibfnamefont{A.~A.} \bibnamefont{Abrikosov}},
  \bibinfo{journal}{Physics Physique Fizika} \textbf{\bibinfo{volume}{2}},
  \bibinfo{pages}{5} (\bibinfo{year}{1965}),
  \urlprefix\url{https://link.aps.org/doi/10.1103/PhysicsPhysiqueFizika.2.5}.

\bibitem[{\citenamefont{Pustilnik and Glazman}(2004)}]{Glazman04}
\bibinfo{author}{\bibfnamefont{M.}~\bibnamefont{Pustilnik}} \bibnamefont{and}
  \bibinfo{author}{\bibfnamefont{L.}~\bibnamefont{Glazman}},
  \bibinfo{journal}{Journal of Physics: Condensed Matter}
  \textbf{\bibinfo{volume}{16}}, \bibinfo{pages}{R513} (\bibinfo{year}{2004}),
  \urlprefix\url{http://stacks.iop.org/0953-8984/16/i=16/a=R01}.

\bibitem[{\citenamefont{Sela et~al.}(2006)\citenamefont{Sela, Oreg, von Oppen,
  and Koch}}]{EranSela}
\bibinfo{author}{\bibfnamefont{E.}~\bibnamefont{Sela}},
  \bibinfo{author}{\bibfnamefont{Y.}~\bibnamefont{Oreg}},
  \bibinfo{author}{\bibfnamefont{F.}~\bibnamefont{von Oppen}},
  \bibnamefont{and} \bibinfo{author}{\bibfnamefont{J.}~\bibnamefont{Koch}},
  \bibinfo{journal}{Phys. Rev. Lett.} \textbf{\bibinfo{volume}{97}},
  \bibinfo{pages}{086601} (\bibinfo{year}{2006}),
  \urlprefix\url{https://link.aps.org/doi/10.1103/PhysRevLett.97.086601}.

\bibitem[{\citenamefont{Nozi\`eres}(1974)}]{Nozieres_FL}
\bibinfo{author}{\bibfnamefont{P.}~\bibnamefont{Nozi\`eres}},
  \bibinfo{journal}{Journal of Low Temperature Physics}
  \textbf{\bibinfo{volume}{17}}, \bibinfo{pages}{31} (\bibinfo{year}{1974}).

\bibitem[{\citenamefont{Jauho et~al.}(1994)\citenamefont{Jauho, Wingreen, and
  Meir}}]{Jauho_1994}
\bibinfo{author}{\bibfnamefont{A.-P.} \bibnamefont{Jauho}},
  \bibinfo{author}{\bibfnamefont{N.~S.} \bibnamefont{Wingreen}},
  \bibnamefont{and} \bibinfo{author}{\bibfnamefont{Y.}~\bibnamefont{Meir}},
  \bibinfo{journal}{Phys. Rev. B} \textbf{\bibinfo{volume}{50}},
  \bibinfo{pages}{5528} (\bibinfo{year}{1994}),
  \urlprefix\url{https://link.aps.org/doi/10.1103/PhysRevB.50.5528}.

\end{thebibliography}

\end{document}